\documentclass[12pt,preprint,showpacs,preprintnumbers,nofootinbib]
{revtex4}

\usepackage{graphicx}% Include figure files
\usepackage{epsfig}
\usepackage{dcolumn}% Align table columns on decimal point
\usepackage{bm}% bold math
\usepackage{amssymb}
\usepackage{here}

\renewcommand{\thefootnote}{\fnsymbol{footnote}}

\def\beq{\begin{equation}}
\def\eeq{\end{equation}}
\def\bea{\begin{array}}
\def\eea{\end{array}}

\def\del{\partial }
\def\to{\rightarrow}

\def\[{\left[}
\def\]{\right]}
\def\({\left(}
\def\){\right)}

\def\slash{\not{}{\mskip-7.mu}}

\def\sm0{{\widetilde{m}_0}}

\def\ov{\overline}

\def\U1em{{U(1)_{\rm em}}}
\def\to{\rightarrow}

\def\sq2{\sqrt{2}}

\def\End{\end{document}}
\def\ra{\rightarrow}
\newcommand{\lae}{\stackrel{<}{\sim}}
\newcommand{\gae}{\stackrel{>}{\sim}}

\def\Journal#1#2#3#4{{#1} {\bf #2} (#4) #3}
\def\NPB{{\em Nucl. Phys.} B}
\def\PLB{{\em Phys. Lett.}  B}
\def\PRL{\em Phys. Rev. Lett.}
\def\PRD{{\em Phys. Rev.} D}
\def\ZPC{{\em Z. Phys.} C}
\def\EPC{{\em Euro. Phys. J.} C}

\begin{document}
% If you do not want PACS numbers comment \draft out.

\title{
Associated Production of CP-odd and Charged Higgs Bosons  \\
at Hadron Colliders}
\author{
{\sc Qing-Hong Cao}$^1$}
\email{cao@pa.msu.edu}
\author{
{\sc Shinya Kanemura}$^2$}
\email{kanemu@het.phys.sci.osaka-u.ac.jp}
\author{
 \, {\rm and} \,
{\sc C.--P. Yuan}$^1$
}
\email{yuan@pa.msu.edu}

\affiliation{
\vspace*{2mm}
{\em $^1$Department of Physics \& Astronomy, Michigan State University,
     East Lansing, Michigan 48824-1116, USA \\
  $^2$Department of Physics, Osaka University,
         Toyonaka, Osaka 560-0043, Japan\\}}
\begin{abstract}

In the Minimal Supersymmetric Standard Model, the masses of the
charged Higgs boson ($H^\pm$) and the CP-odd scalar ($A$) are
related by $M_{H^+}^2=M_A^2+m_W^2$ at the Born level.
Because the coupling of $W^-$-$A$-$H^+$ is fixed by gauge interaction,
the Born level production rate of
$q \bar q' \to W^{\pm \ast} \to A H^\pm$ depends only on
one supersymmetry parameter -- the mass ($M_A^{}$) of $A$.
We examine the sensitivity of the LHC to this signal event
in the $A(\ra b {\bar b})H^+(\ra \tau^+ \nu_\tau)$
and $A(\ra b {\bar b})H^+(\ra t {\bar b})$ decay channels.
We illustrate how to test the mass relation
between $A$ and $H^+$ in case that the signal is found.
If the signal is not found, the product of the decay branching
ratios of $A$ and $H^\pm$ predicted by the MSSM
is bounded from above as a function of $M_A$.

\end{abstract}


\maketitle

\setcounter{footnote}{0}
\renewcommand{\thefootnote}{\arabic{footnote}}

%\def\doublespaced{\baselineskip=\normalbaselineskip\multiply\baselineskip
%  by 150\divide\baselineskip by 100}
%\def\doublespaced{\baselineskip=\normalbaselineskip\multiply\baselineskip
%  by 120\divide\baselineskip by 100}
%\doublespaced

\section{Introduction}

One of the top priorities of current and future high-energy
colliders, such as the Fermilab Tevatron and CERN Large Hadron
Collider (LHC), is to probe the mechanism of the electroweak
symmetry breaking. In the Standard Model (SM) of particle physics,
this amounts to searching for the yet-to-be-found Higgs boson. It
is also possible that the mechanism of electroweak symmetry
breaking originates from new physics beyond the SM. Very often,
the low energy effective theory of new physics model predicts the
existence of extended Higgs sector that contains more than one
Higgs boson. Hence, in order to discriminate each new physics
model from its alternatives, it is important to detect these
additional Higgs bosons and to measure their properties.

Supersymmetry (SUSY) is one of the most commonly studied new
physics models. The Higgs sector of the Minimal Supersymmetric
Standard Model (MSSM) is known as a special case of the Two Higgs
Doublet Model (THDM) with the type-II Yukawa
interaction~\cite{hhg}. The THDM yields five physical scalar
states, i.e., two CP-even ($h$ nd $H$), a CP-odd ($A$) and a pair
of charged Higgs bosons ($H^\pm$).

In contrast to the free parameters in the Higgs sector of
the THDM, the masses and mixing
angles among the Higgs bosons in the MSSM are determined by the
gauge couplings due to the requirement of supersymmetry.
One of the striking features of the MSSM Higgs sector is that
the mass ($M_h$) of the lightest Higgs boson $h$
is predicted to be
 smaller than the mass of the $Z$ boson at the tree level
and less than about 130 GeV after including contribution from
radiative corrections~\cite{h0mass}. The discovery of such a light
Higgs boson could be a strong hint of the MSSM, as the mass of the
SM Higgs boson has to be approximately between 130\,GeV and
180\,GeV for the SM to be a valid theory all the way up to the
Planck scale ($10^{19}$ GeV)~\cite{smh0,quiros}. However, the
discovery of a light Higgs boson does not exclude new physics
models in which the light Higgs boson naturally has a mass less than
130 GeV, even when the Planck scale is taken as the cut-off scale
of the models. Two examples are the non-SUSY THDM and Zee
model~\cite{2hdm}. Therefore, a detailed study of the Higgs sector
is necessary to discriminate the MSSM from other new physics
models. Needless to say that a general test of the MSSM should
also include the experimental identification of super-partner
particles, such as sfermions, charginos and neutralinos.

Many studies have been done in the literature to show how to
detect heavier MSSM Higgs bosons produced at the LHC~\cite{tdr,LH}.
A light charged Higgs boson with mass $M_{H^\pm}^{} < m_t - m_b$ 
may be produced via the top quark decay~\cite{tdecay} 
or via the pair production
$qq \to H^+H^-$~\cite{qqHpHm}, 
$gg \to H^+H^-$~\cite{ggHpHm} and
$qq \to H^+H^-qq$~\cite{qqHpHmqq}.
The main production channels for the heavier $H^{\pm}$ may be
those associated with heavy quarks such as
$g b \to H^- t$~\cite{gb} and $q b \to q' b H^\pm$~\cite{qb}.
The associated production of the $W$ boson and charged Higgs boson 
$b \bar b, \/ gg \to H^\pm W^\mp$~\cite{HpWm}
may also be substantial, especially for small and large
$\tan\beta$ values.
The possible discovery channels from the decay of $H^\pm$ are
$H^\pm \to t b$~\cite{tb}, 
$H^\pm \to \tau\nu$~\cite{taunu}, 
and $H^\pm \to W^\pm h$~\cite{wh}.
Furthermore, the CP-odd Higgs boson may be produced mainly in
association with bottom quarks 
($gg,qq \to A b {\bar b}$)~\cite{Abb}. 
The possible discovery channels from the decay
of $A$ are $A \to t \bar t$~\cite{LH}, $A \to \gamma\gamma$~\cite{Agamgam} 
and $A \to Z h$~\cite{Zh0} for low $\tan\beta$ values, and $A \to
\mu^+\mu^-$~\cite{mumu} or $\tau\bar\tau$~\cite{tautau} for large
$\tan\beta$ values. The production and decay channels discussed
above can be used to probe a large part of MSSM parameter space in
the $M_A^{}$ versus $\tan\beta$ plane, except for intermediate
values of $\tan\beta$ and large values of $M_A$ where only the
lightest Higgs boson $h$ is likely to be found.

In a recent paper \cite{wah_plb}, two of us proposed to
study the associated production of CP-odd scalar ($A$) and
charged Higgs boson ($H^\pm$) via $q \bar q' \to W^{\pm \ast} \to A H^\pm$
at hadron colliders to test the MSSM. It was shown that at the Born level
the production of this process depends
only on one SUSY parameter --
the mass ($M_A^{}$) of $A$. This is because in the MSSM the
masses of $H^\pm$ and $A$ are related by the mass of the
$W^\pm$ boson ($m_W$) as\footnote{
We note that in general, a CP
violating phase can enter the Higgs sector of the MSSM, so that
the CP-even Higgs bosons can mix with the CP-odd
Higgs scalar and the mass relation~(\ref{eq:massrel}) does
not hold any more. In this paper, we shall focus our study on the
MSSM with a CP invariant Higgs sector.
}
\begin{eqnarray}
M^2_{H^\pm} = M^2_A + m^2_W \, ,
\label{eq:massrel}
\end{eqnarray}
and the coupling of
$W^-$-$A$-$H^+$ is fixed by gauge interaction.
In case that $M_A$ can be measured from data by
reconstructing its decay particles
(such as $b {\bar b}$ pair),
the Born level production rate of this process
is uniquely determined and
does not depend on any other SUSY parameters.
Therefore, studying this process can test
the product
of the Higgs boson decay branching ratios (in contrast to
the product of decay branching ratios {\it and} production rate)
predicted by the MSSM.
The above conclusion holds for any SM
decay model of $A$ and $H^\pm$.
For example, if the decay mode of
$A \to b \bar b$ and $H^+ \to \tau^+ \nu_\tau$
in the $AH^+$ event is studied and no excess is found in 
 experimental data for a given mass bin of
$M_A$ (hence, $M_{H^\pm}$), one can constrain
the MSSM by demanding the product of
the decay branching ratios, ${\rm Br}(A \to b {\bar b}) \times {\rm
Br}(H^+ \to \tau^+ \nu_\tau)$, to be bounded from above as a
function of $M_A$. Similarly, applying the same strategy, one
can constrain the product
${\rm Br}(A \to X) \times {\rm Br}(H^+ \to Y)$
for any decay modes $X$ and $Y$ predicted by the MSSM,
as a function of only one SUSY parameter -- $M_A$.

Another interesting feature of this process is that the kinematic
acceptance (therefore, the detection efficiency) of the signal
events does not depend on the choice of the other SUSY parameters
because both $A$ and $H^\pm$ are spin-0 (pseudo-)scalar particles
so that the kinematic distributions of their decay particles can
be accurately modelled. (It is an isotropic distribution in the
rest frame of $A$ or $H^\pm$.) Moreover, the one-loop
electroweak corrections to the mass relation~(\ref{eq:massrel})
and the $W^-AH^+$ coupling are generally smaller than the other
theoretical errors (such as the parton distribution function
uncertainties) and the expected experimental errors (such as the
mass resolution of Higgs boson decaying into jets). 
Despite that for some
extreme choice of the MSSM parameters, the mass
relation~(\ref{eq:massrel}) can receive a relatively large
correction~\cite{akeroyd}, its effect to the $AH^+$ production
rate is found to be small. Therefore, in this work  
we shall only perform a Born level Monte Carlo analysis to 
study the sensitivity of the LHC to this production channel.

To detect the signal event at hadron colliders,
it is necessary to suppress the potentially large background
rates predicted by the SM.
In this paper, we show how to detect such a signal
at the LHC for the specific
decay mode of $AH^\pm \to b {\bar b} \tau^+ \nu_\tau$
with $\tau^+ \to \pi^+ {\bar \nu}_\tau$.
While it is possible to include other decay channels of $AH^\pm$,
we shall restrict ourselves to this simple decay mode and use
it as an example to illustrate the idea of Ref.~\cite{wah_plb}.
Namely, studying the associated production of $A$ and $H^\pm$ at
hadron colliders can discriminate the MSSM from its alternatives,
e.g., a general two-Higgs-doublet model (THDM), and
test the prediction of the MSSM on the product
of the decay branching ratios of $A$ and $H^\pm$.

The rest of the paper is organized as follows. In Sec. II, we
briefly review the result of Ref.~\cite{wah_plb} for the
production rate of the signal event at the Fermilab Tevatron and
CERN LHC
 up to the one-loop
order QCD and electroweak corrections.
In Sec. III, we present a Monte Carlo analysis for studying the
sensitivity of the LHC to the detection of the signal event.
We show that all the dominant SM backgrounds can
be effectively suppressed by making use of the polarization of
$\tau^\pm$.
Sec. IV contains our conclusions.

\section{Production Cross Section of Signal}
\begin{figure}[t]
\begin{center}
\scalebox{0.6}{\includegraphics{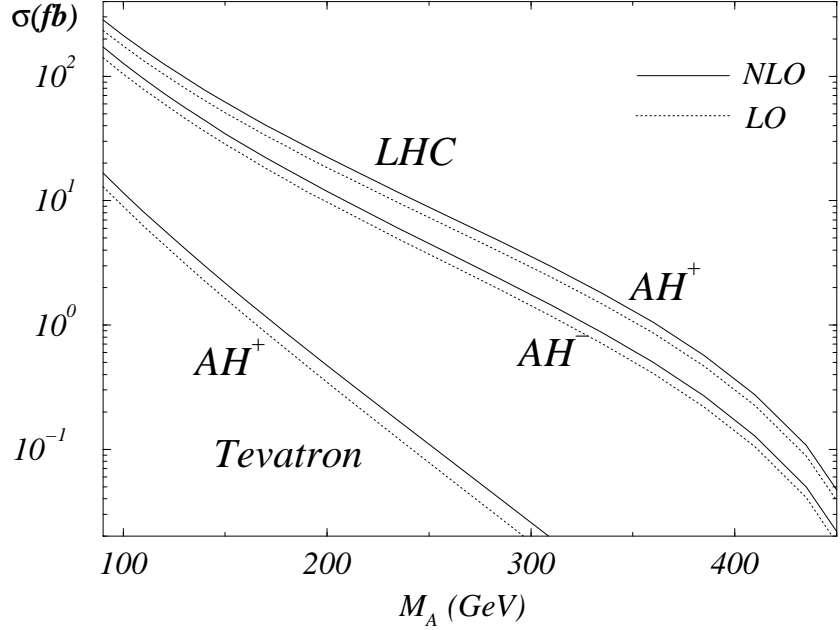}}
\end{center}
\vspace*{-5mm}
\caption{The LO (dotted lines) and NLO QCD (solid lines)
cross sections of the $AH^+$ and $AH^-$ pairs
as a function of $M_A^{}$ at the Tevatron (a 1.96\,TeV $p \bar p$ collider),
and the LHC (a 14\,TeV $p p$ collider).
The cross sections for $AH^+$
and $AH^-$ pair productions coincide at the
Tevatron for being a $p \bar p$ collider. }
\label{fig:cross}
\end{figure}

At hadron colliders, the Born level process
for the production of a $A H^\pm$ pair is
$q {\bar q'} \to W^{\pm\ast} \to A H^\pm$.
The coupling of
$W^\mp$-$A$-$H^\pm$\ is induced from the gauge invariant
kinetic term of the Higgs sector \cite{hhg}:
\begin{eqnarray}
{\cal L}_{int}= \frac{g}{2} W^+_\mu (A \del^\mu H^- - H^- \del^\mu A)
+ {\rm h.c.} \, ,
\end{eqnarray}
and its strength is determined by the
weak gauge coupling $g$.
Hence, the Born level cross section
$\sigma(p {\bar p}, \, pp \to A H^\pm)$
depends only on the masses of $A$ and $H^\pm$.
Its squared amplitude, after averaging over the spins and colors is
\begin{eqnarray}
{\overline {|{\cal M}|^2} } = {4 \over 3} m_W^4 G_F^2
{ {s} \over ({s}^2 -m_W^2)^2 + m_W^2 \Gamma_W^2}
P^2 \sin^2\theta \, ,
\end{eqnarray}
where $P=\sqrt{E_A^2-M_A^2}\, $ with
$E_A=({s}+M_A^2-M_{H^+}^2)/(2 \sqrt{{s}}) \, $
and
$\theta$ is the polar angle of $A$ in the center-of-mass (c.m.) frame
of $A$ and $H^\pm$.
In the MSSM, the
masses of the charged Higgs boson $H^\pm$
and the CP-odd scalar $A$ are strongly correlated.
At the Born level,
$M_{H^\pm} = \sqrt{M^2_A + m^2_W}$.
In Fig.~\ref{fig:cross}, we show the inclusive production
rate of $A H^\pm$ as a function of
$M_A$ for the Tevatron (a 1.96\,TeV proton-antiproton collider)
and the LHC (a 14\,TeV proton-proton collider).
Here, the CTEQ6M parton distribution
functions (PDF) \cite{cteq6} are used and both the renormalization
and the factorization scales are chosen to be the
invariant mass ($\sqrt{s}$) of the $A H^\pm$ pair.
We note that for the $c \bar b \to A H^+$ subprocess,
 in addition to the Cabbibo-Kobayashi-Maskawa
suppressed $s$-channel $W$-boson diagram, there is
a $t$-channel diagram whose contribution becomes large
for a large value of $\tan \beta$.
Nevertheless, the $c \bar b \to A H^+$ contribution
to the inclusive $AH^+$ rate is small. For example, its contribution
to the total rate is less than 0.01\% and 0.1\% at the
Tevatron and the LHC, respectively,
for $\tan \beta=40$ and $M_A=90$\,GeV.
For a smaller value of $\tan\beta$, its contribution becomes negligible.
Hence, we shall ignore its contribution in the following discussion.

It is easy to include the
next-to-leading order (NLO) QCD correction to calculate the
production rate of the signal event because at the one-loop order only
the initial state receives radiative contribution which is
similar to the NLO QCD correction to the $W$-boson
production, but with a different invariant mass, at hadron colliders.
The result is also shown in Fig.~\ref{fig:cross}.
We find that typically the NLO QCD production rate is about
20\% higher than the leading-order (LO) rate when the same PDF is
used. The higher order ($\alpha^2_s$
or above) QCD correction is estimated to be about 10\% at
the Tevatron and less than a percent at the LHC, for $M_{A}^{}=120$
GeV, when the factorization scale is varied around
the $A H^\pm$ invariant mass $\sqrt{{s}}$ by a factor of 2.

The dominant one-loop electroweak corrections to the
$q{\bar q'} \to A H^\pm$ process come from the loops of top ($t$) and
bottom ($b$) quarks as well as their supersymmetric partners, i.e.
stops ($\tilde{t}_{1,2}$) and sbottoms ($\tilde{b}_{1,2}$), in the
MSSM, due to their potentially large couplings to Higgs bosons.

In the following, we discuss the quark- and squark-loop
radiative corrections to the effective coupling of $W^\pm A H^\mp$ and
to the mass relation~(\ref{eq:massrel}).
In our calculation, we adopt the on-shell renormalization scheme
developed by Dabelstein in Ref.~\cite{dabelstein}.
(For completeness, we outline the renormalization scheme
in Appendix {\bf A}).

The part of the one-loop effective coupling of
$W^\pm A H^\mp$, that is relevant to the production
process $q {\bar q'} \to H^+A$, can be written as\footnote{
The other form factor, $(p_{A}^{} + p_{H}^{})^\mu$,
does not contribute to this process for massless quarks.
}
%\vspace*{-2mm}
\noindent
\begin{eqnarray}
  M_{WHA}^\mu(q^2) &=&
 -\frac{{\bar g}}{2} (p_{A}^{} - p_{H}^{})^\mu \,
  \left[ 1 + F^{(1)}(q^2) \right],
\end{eqnarray}
%
%\vspace*{-2mm}
\noindent
where  $q^\mu$, $p_A^\mu$ and
$p_H^\mu$ are the incoming momenta of
$W^+$, $A$ and $H^-$, respectively,
and ${\bar g}$ is the effective weak gauge coupling evaluated
at $q^2$.
Hence, the radiative correction to the cross section of the sub-process
$q {\bar q'}\to AH^+$ at the one-loop order is
\begin{eqnarray}
  K^{(1)}(q^2)  \equiv 2 {\rm Re}\, F^{(1)}(q^2) \, ,
\label{eq:kfactor}
\end{eqnarray}
where the detailed calculation for $F^{(1)}(q^2)$ is summarized in
Appendix~{\bf B}.
As shown in Eqs.~(\ref{eq:qform1}), (\ref{eq:qform2}) and (\ref{eq:qform3}),
the quark-loop contribution is proportional to
the squared Yukawa coupling constants
$y_t^2 (= 2 m_t^2 \cot^2\beta/v^2)$ and
$y_b^2 (= 2 m_b^2 \tan^2\beta/v^2)$.
In the large $m_t$ or large $m_b \tan\beta$ limit,
it can be written as
%
%\vspace*{-2mm}
\noindent
\begin{eqnarray}
 F^{(1)}_{\rm quark}
 &\sim& \frac{N_c}{16\pi^2}
 \left[ -\frac{1}{4} y_t^2
  +  \frac{1}{2} \left( \frac{3}{2} - \ln \frac{m_t^2}{m_b^2} \right)
        y_b^2  \right],
\end{eqnarray}
%
%\vspace*{-2mm}
\noindent
where $N_c(=3)$ is the number of colors.
Since $y_t^2$ and $y_b^2$ are at most ${\cal O}(1)$
for $\tan\beta\simeq 1$ and $m_t/m_b$, respectively,
$F^{(1)}_{\rm quark}$ is at most a few percent for
$1 \lae \tan\beta \lae m_t/m_b$.

We also calculate the squark-loop contribution.
As compared to the quark-loop effects,
the squark-loop effects are rather complex
because of the additional SUSY parameter dependence.
The mass eigenstates $\tilde{f}_{1,2}$
($\tilde{f}=\tilde{t}$ or $\tilde{b}$)
of the squarks
are obtained from the weak eigenstates
$\tilde{f}_{L,R}$ by diagonalizing the mass matrices
defined through \cite{hk}
%
%\vspace*{-2mm}
\noindent
\begin{eqnarray}
  {\cal L}_{\rm mass} = - (\tilde{f}_L^\ast, \tilde{f}_R^\ast)
        \left( \begin{array}{cc}
               M_L^2 & m_{f} X_{f}\\
               m_{f} X_{f} & M_R^2\\
         \end{array} \right)
        \left( \begin{array}{c}
              \tilde{f}_L \\ \tilde{f}_R
         \end{array} \right),
\end{eqnarray}
\noindent
where, $M_L^2=M_{\tilde{Q}}^2+m_f^2+
       (m_Z^2 \cos 2\beta) (T_{f_L^{}}-Q_f  s_W^2)$ and
       $M_R^2=M_{\tilde{U}, \tilde{D}}^2+m_f^2+
       (m_Z^2 \cos 2\beta) Q_f s_W^2$.
In this expression,
$M_{\tilde{Q}}^2$, $M_{\tilde{U}}^2$ (for $\tilde{f}=\tilde{t}$)
and $M_{\tilde{D}}^2,$ (for $\tilde{f}=\tilde{b}$)
are the soft-breaking masses for
$\tilde{f}_L$, $\tilde{t}_R$ and $\tilde{b}_R$, respectively;
$s_W=\sin \theta_W$ with $\theta_W$ being the weak mixing angle;
$T_{f_L^{}}$ is the isospin of the left-handed quark $f_L$;
and $Q_f$ is the electric charge of the quark $f$.
Moreover,
$X_t =   A_t - \mu \cot\beta$ and
  $X_b =   A_b - \mu \tan\beta$,
where $A_t$ ($A_b$) is the trilinear $A$-term for
${\tilde t}$ (${\tilde b}$), and
$\mu$ is the SUSY invariant higgsino mass parameter \cite{hk}.
For completeness, we have listed all the relevant squark and Higgs boson
couplings in {Appendix~{\bf C}}, so that
the squark-loop contributions to $F^{(1)}(s)$,
cf. Eq.~(\ref{eq:ff}), can be directly read
out from the results in
Eqs.~(\ref{eq:sqform1}), (\ref{eq:sqform2}) and (\ref{eq:sqform3}).

To examine the effect of one-loop electroweak corrections,
we shall discuss two limiting cases below.
Firstly, we consider the cases with $\mu=A_t=A_b=0$, i.e., the
cases without stop mixing ($|X_t|=0$) and sbottom mixing
($|X_b|=0$). Under this scenario, the masses of squarks are
proportional to the typical scale of the soft-breaking mass,
denoted as $M$. Since all the relevant couplings between squarks
and Higgs bosons are independent of the soft-breaking masses
$M_{\tilde{Q}}$, $M_{\tilde{U}}$ and $M_{\tilde{D}}$ (see
Appendix~{\bf C}), the squark-loop effect is decoupled and its
contribution is very small for a large value of $M$, where $M
\equiv M_{\tilde{Q}} \simeq M_{\tilde{U}} \simeq M_{\tilde{D}}$.
 For a smaller $M$, $F^{(1)}_{\rm squark}$
becomes larger. However, $M$ cannot be too small because a small
$M$ implies light squarks whose masses are already bounded from
below by the direct search results \cite{PDG}. Furthermore, as to
be shown later, the case with a small $M$ is also strongly
constrained by the $\rho$ parameter measurement.
Secondly,
we examine the case with a large stop mixing, assuming
$m_t |X_t| \sim M^2 \gg m_Z^2$.
Such a large stop mixing leads to a large mass splitting between
$\tilde{t}_1$ and $\tilde{t}_2$ so that
$m _{\tilde{t}_1} \simeq {\cal O}(m_Z)$ and
$m_{\tilde{t}_2} \simeq \sqrt{2} M$, while
$m_{\tilde{b}_{1,2}} \simeq M$.
The leading squark contribution to $F^{(1)}(q^2)$
can be expressed as
%
%\vspace*{-3mm}
\begin{eqnarray}
\label{eq:fsquark}
\!\!\!\!\!\!\!\!\!\!\!\!\!\!\!\!\!\!\!\!
\!\!\!\!\!\!\!\!\!\!\!\!\!\!\!\!\!\!\!\!
  F^{(1)}_{\rm squark}
 \sim \frac{-N_c}{16\pi^2}
 \left[  \left(\frac{3}{4}-\ln{2} \right)
    \left(\frac{Y_{\tilde t}}{M}\right)^2
+ \left( \frac{13}{6} -3 \ln{2} \right)
      \left(\frac{Y_{\tilde b}}{M} \right)^2  \right],
\end{eqnarray}
%
%\vspace*{-3mm}
%\noindent
with
 $Y_{\tilde{t}} =  \frac{m_t}{v} (A_t \cot\beta+\mu)$ and
 $Y_{\tilde{b}} =  \frac{m_b}{v} (A_b \tan\beta+\mu)$.
Since in this case $|A_t| \simeq |M^2/m_t \pm |\mu| \cot\beta|$,
we have $|Y_{\tilde{t}}|
\lae {\cal O} (M^2/v)$
for $|\mu| \lae M$ and $\tan\beta \gae 1$.
When $|A_b| \simeq |A_t| > |\mu|$ and $\tan\beta \lae m_t/m_b$,
we find $|Y_{\tilde{b}}| \lae {\cal O} (M^2/v)$.
Thus, with a large stop mixing ($m_t |X_t| \simeq M^2$),
$F^{(1)}_{\rm squark}$ is proportional to the soft-breaking
mass scale $M$, and does not decouple in the large $M$ limit.
However, such kind of model is strongly
constrained by the $\rho$-parameter (or the $T$-parameter) data.
With a large stop mixing ($M^2 \simeq m_t |X_t|$),
the squark contribution to the $\rho$-parameter (cf. Appendix~{\bf D})
is
%
%\vspace*{-2mm}
\begin{eqnarray}
 \Delta \rho_{\rm squark}
 \simeq
(2.2\times 10^{-3}) \frac{M^2}{v^2}.  \label{eq:rho}
\end{eqnarray}
%
%\vspace*{-2mm}
%\noindent
Since any new physics contribution to the $\rho$-parameter
has to be bounded by data as \cite{rho}
%
%\vspace*{-2mm}
%\noindent
\begin{eqnarray}
  -1.7 < \Delta  \rho_{\rm new} \times 10^3  < 2.7,
  \;\;\;\; {\rm at} \,2\sigma \,{\rm level,}  \label{rho}
\end{eqnarray}
%
%\vspace*{-2mm}
%\noindent
the scale $M$ cannot be too large in this case.
Consequently, the above $F^{(1)}_{\rm squark}$
is constrained to be
smaller than a few percent as long
as  $\mu^2$ is not much larger than $M^2$.

\begin{figure}[t]
\begin{center}
\scalebox{0.6}{\includegraphics{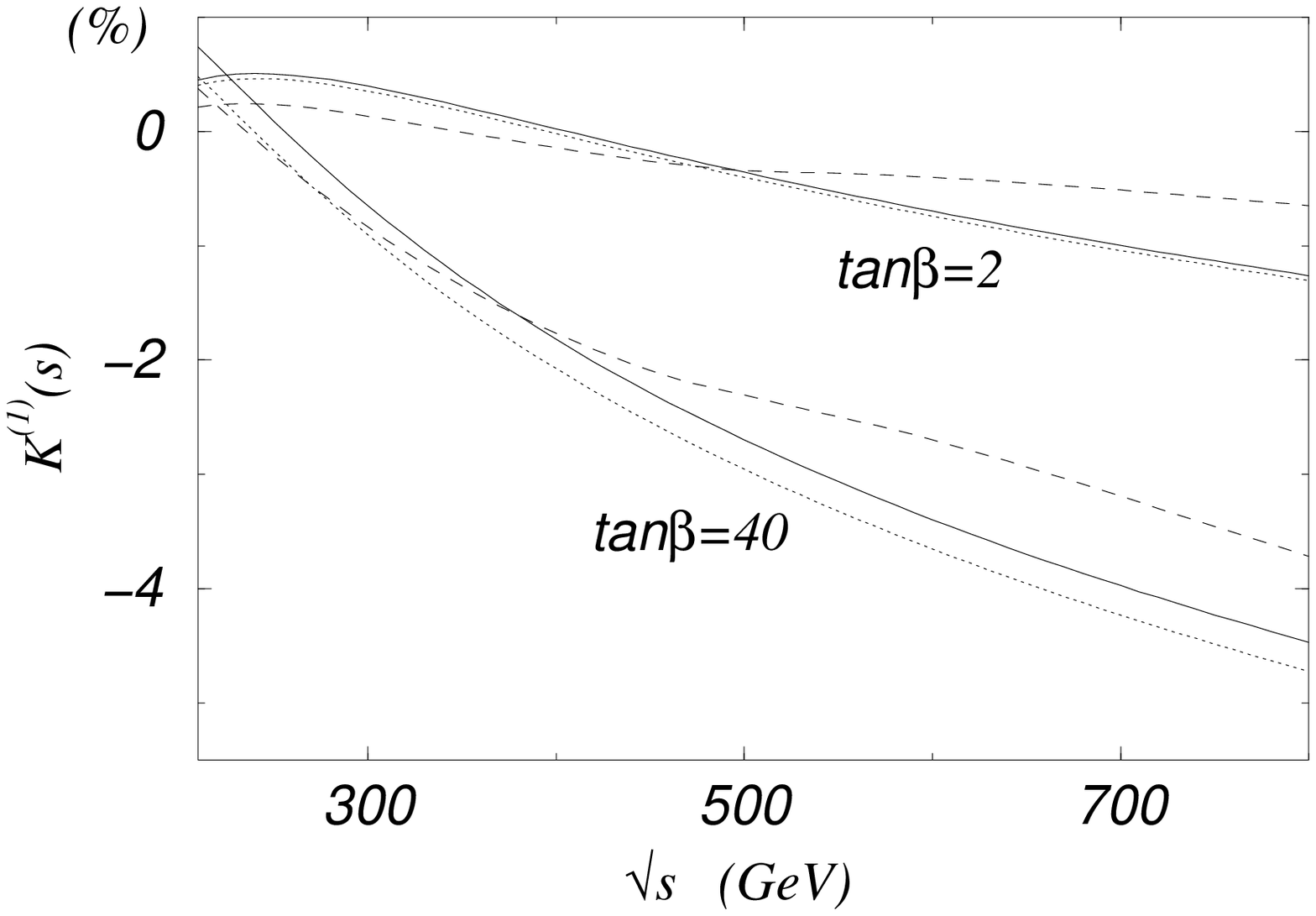}}
\end{center}
\vspace*{-5mm}
\caption{The $K$-factor, $K^{(1)}(s)$, of the
constituent process $q {\bar q'} \to H^+A$ for $M_A^{}=90$ GeV,
as a function of the invariant mass $\sqrt{s}$ of $q {\bar q'}$.
The solid lines correspond to the top and bottom quark contribution.
The cases where the squark-loop contribution is included are
described by the dotted lines
for those without stop mixing (Set 1 and Set 2) and by dashed lines
for those with maximal stop mixing (Set 3 and Set 4), respectively. }
\label{fig:kfactor}
\end{figure}

To examine the effect from the stop and sbottom
loops to the production rate of $AH^\pm$, we consider 4 sets of
SUSY parameters, as listed in Table \ref{tbl:susypara1},
which give the largest allowed deviation in the
$\rho$-parameter.
Set~1 and Set~2 represent the cases without either a stop
mixing ($X_t=0$) or a sbottom mixing ($X_b=0$), and
Set~3 and Set~4 are the cases with a large stop
mixing ($m_t |X_t| \simeq M^2$) and $m_{\tilde{t}_1}\simeq
100$ GeV.
The $K^{(1)}(s)$ factor,
as defined in Eq.~(\ref{eq:kfactor}),  is shown
in Fig.~\ref{fig:kfactor} as a function of the
invariant mass ($\sqrt{s}$) of
the constituent process for $M_A^{}=90$ GeV.
It is clear that the quark-loop contribution to $K^{(1)}(s)$
dominates the squark-loop contribution.
For these four sets of SUSY parameters,
the squark-loop contribution is smaller than
the quark-loop contribution
by about a factor of 100.
Generally, the squark contributions are at most a few percent,
unless $|\mu|$ is taken to be very large as compared to
the scale $M$.
We have checked that
this conclusion does not change when the
assumption $M_Q^2 \simeq M_U^2 \simeq M_D^2$
is relaxed to some extent.
Including both the quark- and squark-loop contributions to $K^{(1)}(s)$,
we found that the correction
to the hadronic cross section of $H^+A$ production
in the invariant mass region just above the $H^+A$ threshold,
where the constituent cross section is the largest, is at
a percent level.
In summary, to be consistent
with the low-energy data
and the direct search results for stops and sbottoms,
the one-loop electroweak correction to the production rate
of $pp, \, p {\bar p} \to A H^\pm$ is small (at most a few percent).

\begin{table}
\caption{The SUSY (input and output) parameters used in
Fig.~\ref{fig:kfactor}.}
\begin{tabular}{c|cccc}
\hline
\hline
 & Set1 & Set2 & Set 3 & Set4  \\ \hline
$M_{\tilde{Q}}=M_{\tilde{U}}=M_{\tilde{D}}$ (GeV)
& 106 & 84 & 408 & 409 \\
$\tan\beta$ &2 &40 &2 &40 \\
$A_t=A_b$ (GeV) & 0 & 0 & $+1261$ & $+1119$ \\
$\mu$ & 0 & 0 & +300 & +300 \\ \hline
\hline
$m_{\tilde{t}_1}$ (GeV) &197 & 184 & 100 & 100\\
$m_{\tilde{t}_2}$ (GeV) &199 & 188 & 612 & 611\\
$m_{\tilde{b}_1}$ (GeV) &108 & 88  & 407 & 373\\
$m_{\tilde{b}_2}$ (GeV) &116 & 103 & 412 & 447 \\ \hline
\hline
$\Delta\rho_{\rm squarks} \times 10^3$   & 2.72 & 2.70 & 2.71 & 2.70 \\
\hline
\hline
\end{tabular}
\label{tbl:susypara1}
\end{table}

Next, we discuss the one-loop corrections to the mass
relation~(\ref{eq:massrel}). Let us parameterize the deviation
from the tree-level relation by $\delta$, so that at the one-loop
order
%
%\vspace*{-2mm}
\begin{eqnarray}
M_{H^\pm}^{} = \sqrt{M_{A}^2 + m_W^2} \left( 1 + \delta \right).
\label{eq:delta}
\end{eqnarray}
%
%\vspace*{-2mm}
%\noindent
We note that in our renormalization scheme (see Appendix~{\bf A}),
$M_A$ and $m_W$ are the input parameters, but $M_{H^+}$ is not.
The one-loop corrected mass of the charged Higgs boson
$M_{H^\pm}^{}$ can be obtained by solving
\begin{eqnarray}
 0 = {\rm Det} \left|
 \begin{array}{cc}
\Gamma^{(2)}_{G^+G^-}(p^2) & \Gamma^{(2)}_{G^+H^-}(p^2) \\
\Gamma^{(2)}_{H^+G^-}(p^2) & \Gamma^{(2)}_{H^+H^-}(p^2) \\
 \end{array} \right|,
\end{eqnarray}
where $\Gamma^{(2)}_{ij}(p^2)$ represent the renormalized
two-point functions in the basis of the renormalized Goldstone
boson ($G^\pm$) and charged Higgs boson ($H^\pm$) fields. Here,
the notation ``${\rm Det}$'' denotes taking the determinant of the
$2 \times 2$ matrix. One of the solutions of the above equation is
$p^2=0$, which corresponds to the charged Nambu-Goldstone mode,
and another is $M_{H^\pm}^2$. At the one-loop level, the pole mass
of the charged-Higgs boson, in our renormalization scheme,
 can be calculated from

\begin{eqnarray}
M_{H^\pm}^2 &=& M_A^2 + m_W^2 \nonumber \\
&&           + \Pi_{AA}(M_A^2) - \Pi_{H^+H^-}(M_A^2+m_W^2)
           + \Pi_{WW}(m_W^2), \label{eq:rmch}
\end{eqnarray}
where
$\Pi_{\phi\phi}(q^2)$ ($\phi=A,H^\pm$,and $W$) are
the self-energies.
For completeness, we summarize
the quark and squark contributions to the self-energies
of $A$ and $H^\pm$ in Appendix~{\bf E}.

When $A_{t,b}$ and $\mu$ are zero (i.e., no-mixing case),
the leading contribution to $\delta$ is found to be
%\vspace*{-0.2cm}
\begin{eqnarray}
\delta &\sim& \frac{N_c}{8\pi^2 v^2}
 \left(\frac{m_t^2 m_b^2}{M_{A}^2+m_W^2}\right)
\frac{1}{\sin^2\beta \cos^2\beta}
\left( 1 +\ln \frac{M^2}{m_t^2} \right).
\end{eqnarray}
This correction is substantial for $\tan\beta \simeq m_t/m_b$ and
$M^2 \gg m_t^2$. Applying Eq.~(\ref{eq:rmch}) with the complete
expression of $\Pi_{\phi\phi}(q^2)$, we found $\delta$ to be less
than 4.9\% for $2 < \tan\beta < 40$, $M < 2$\,TeV and $M_A >
90$ GeV. Our result agrees well with Ref.~\cite{masses1}, in which
an approximate formula was presented for $M^2 \gg m_t^2$.

For the cases with nonzero $A_{b,t}$ and $\mu$ parameters,
$\delta$ receives extra contributions, which are
proportional to $A_{t,b}^4/M^4$, $A_{t,b}^2 \mu^2/M^4$
and $\mu^4/M^4$~\cite{masses2,masses1} originated from the squark
couplings [cf.
Eqs.~(\ref{coup1}), (\ref{coup2}), (\ref{coup3}) and (\ref{coup4})]
and squark masses.
For the Set 3 and Set 4 parameters listed in Table~\ref{tbl:susypara1},
$\delta$ is found to be
less than 5.3\% and 3.6\% for $M_A > 90$ GeV, respectively.
In summary, as long as $|A_{t,b}|$ and $|\mu|$ are not too large
as compared to $M$, in a wide range of the parameter space
that is allowed by the
available experimental and theoretical constraints,
$\delta$ does not exceed 10\%.
(In some corner of SUSY parameter space,
$\delta$ can be larger than 10\%~\cite{akeroyd}.
Nevertheless, the $AH^+$ production rate at the LHC
will not be strongly modified as long as $M_A$ is not too
large so that the mass threshold effect becomes important.)

Supported by the finding that
the one-loop electroweak corrections to the $W^\pm A H^\mp$ coupling
and to the mass relation $M^2_{H^\pm}=M_{A}^2+m_W^2$
are generally smaller than the other theoretical errors
(such as the parton distribution function uncertainties)
and the expected experimental errors
(such as the mass resolution of Higgs boson decaying
into jets, which is about 10\,GeV for a 100\,GeV Higgs boson),
we conclude that studying the process
 $ p {\bar p} \, , pp \to W^{\pm \ast} \to A H^\pm$
allows us to distinguish the MSSM from its alternatives by verifying
the mass relation~(\ref{eq:massrel}), or
Eq.~(\ref{eq:delta}), when the signal is found.
If the signal is not found, studying this process
 can provide an upper bound
on the product of the decay branching ratios of $A$ and $H^\pm$ as
a function of only one SUSY parameter -- $M_A$.

\section{Monte Carlo Analysis}

In this section we discuss how to detect such a signal
at hadron colliders.
As a concrete example, we examine the specific
decay channel of $AH^\pm \to b {\bar b} \tau^+ \nu_\tau$
with $\tau^+ \to \pi^+ {\bar \nu}_\tau$.
As shown in Fig.~\ref{fig:cross},
the signal cross section at the Tevatron is much smaller than that at
the LHC. Hence, we shall first discuss its phenomenology at the LHC,
and then comment on the result at the Tevatron.
Furthermore, since the NLO QCD and electroweak corrections to the signal
cross section are small, for simplicity, we shall
perform the Monte Carlo analysis at the Born level.

In the MSSM, the mass of the heavier CP-even Higgs boson ($H$) is
about the same (approximately within 10\,GeV)
as $M_A$ for $M_A \gae 120$ GeV and $\tan\beta \gae
10$. In this case, $q{\bar q'} \to H H^\pm$ can produce similar final
states as $q{\bar q'} \to A H^\pm$.
In contrast to the coupling of
$W^\pm A H^\mp$ whose strength is $\displaystyle \frac{g}{2}$, the tree level
coupling of $W^\pm H H^\mp$ has the strength of
$\displaystyle \frac{g}{2} \sin(\beta-\alpha)$,
though with the same momentum dependence.
Since $\sin^2(\beta-\alpha)\simeq 1$
for $M_A \gae 190$ GeV and $\tan\beta \gae 10$,
the production rate of $HH^\pm$ is almost the same as $AH^\pm$
in the decoupling limit.
Thus, for a large $M_A$,
when both of them decay into the same decay channels,
it will be very difficult to separate the production of $AH^\pm$ from
$H H^\pm$ unless a fine mass resolution can be achieved experimentally.
Nevertheless, studying different decay channels can help to
separate these two production modes. For example, a heavy
$H$ can decay into
a $ZZ$ pair at the Born level, but $A$ cannot.
In the following analysis, we shall consider both the
$AH^\pm$ and $H H^\pm$ production channels
that contribute to the same final state.

Either in the MSSM or the Type-II THDM,
for a large $\tan \beta$ value,
the dominant decay mode is $h,H,A \rightarrow b\ov{b}$
for the neutral Higgs bosons ($h,H,A$),
and
$ H^+ \rightarrow \tau^+ \nu $
for a charged Higgs boson whose mass is smaller than the 
top quark mass.
Due to the missing energy carried away by the
final state neutrino,
it is not possible to directly reconstruct the mass of $H^+$
in the decay mode $ H^+ \rightarrow \tau^+ \nu $.
But, the transverse mass of the charged Higgs boson
can be reconstructed from the $\tau$ jet momentum
and missing transverse energy~\cite{taunu}.
For $m_{H^{\pm}}\gtrsim 200$\,GeV, the dominant decay mode is
$H^{+}\rightarrow t \ov{b}$, in which $M_{H^+}$ can be reconstructed
after properly choosing the longitudinal momentum of the neutrino
(with a two-fold solution) from $t$ decay.
To cover both decay modes in our study,
we consider the following three benchmark
cases with $\tan \beta = 40$:
$(i)$ $M_A=101$\,GeV (and  $M_{H^+} < m_t + m_b$), and
$H^{+}\rightarrow \tau \nu$ being the dominant decay mode;
$(ii)$ $M_A=166$\,GeV (and  $M_{H^+} \sim m_t + m_b$),
and
$H^{+}\rightarrow \tau \nu$ being the dominant decay mode;
$(iii)$ $M_A=250$\,GeV and
$H^{+}\rightarrow t\ov{b}$ being the dominant decay mode.

\begin{table}
\caption{The mass, width,
production rate and decay branching ratio of Higgs bosons
 used in the Monte Carlo analysis. Here,
 $\tan \beta=40$, and all the other SUSY parameters
(soft-breaking masses and $\mu$-parameter) 
 are taken to be 500\,GeV.}
\begin{tabular}{c|c|c|c}
\hline
Sets&
\multicolumn{1}{c|}{A} & B & C \tabularnewline
\hline
\hline
$m_{A}/\Gamma_{A}$\,(GeV)& 101 / 3.7 & 166 / 5.6 & 250 / 7.9 \tabularnewline
\hline
$m_{h}/\Gamma_{h}$\,(GeV)&97 / 3.3 & 112 / 0.04 & 112 / 0.01 \tabularnewline
\hline
$m_{H}/\Gamma_{H}$\,(GeV)&113 / 0.38 & 163 / 5.5  & 248 / 7.8\tabularnewline
\hline
$m_{H^{+}}/\Gamma_{H^{+}}$\,(GeV)& 126 / 0.43 & 182 / 0.68 & 261 / 4.2
\tabularnewline
\hline
$\sigma(AH^{+})$ $fb$ & 164 & 36.5 & 5.4\tabularnewline
\hline
$\sigma(HH^{+})$ $fb$ & 15.0 & 37.2 & 5.4\tabularnewline
\hline
$B(A\rightarrow b\bar{b})$& 0.91 & 0.90 & 0.89\tabularnewline
\hline
$B(H\rightarrow b\bar{b})$& 0.90 & 0.90 & 0.89\tabularnewline
\hline
$B(H^{+}\rightarrow\tau^{+}\nu)$& 0.98& 0.90& 0.21\tabularnewline
\hline
$B(\tau^{+}\rightarrow\pi^{+}\nu)$& 0.11& 0.11& 0.11\tabularnewline
\hline
$B(H^{+}\rightarrow t\bar{b})$& 0.00& 0.09& 0.79\tabularnewline
\hline
$B(t\rightarrow b e^+ \nu)$& 0.11& 0.11&0.11\tabularnewline
\hline
$\sin(\beta-\alpha)$ & 0.33 & 0.997 & 0.999 \tabularnewline
\hline
\hline
\end{tabular}
\label{tbl:susypara2}
\end{table}

\subsection{$M_A=101$\,GeV}

Taking $M_A=101$\,GeV, $\tan \beta=40$,
and all the other SUSY parameters
(soft-breaking masses and $\mu$-parameter) to be 500\,GeV,
one can calculate the masses, widths and decay branching ratios
of the Higgs bosons
using the numerical program HDECAY~\cite{hdecay}.
We find that $M_h=97$\,GeV,
$M_H=113$\,GeV and $M_{H^\pm}=126$\,GeV, and
the relevant decay branching ratios are $B(A \to b {\bar b})=0.91$,
$B(H \to b {\bar b})=0.90$, $B(H^+ \to \tau^+ {\nu})=0.98$,
and $B(\tau^+ \to \pi^+ {\bar \nu})=0.11$.
The relevant total decay widths are
$\Gamma_A=3.7$\,GeV, $\Gamma_H=0.38$\,GeV and
$\Gamma_{H^\pm}=0.43$\,GeV.
For clarity, we summarize the above results in Table~\ref{tbl:susypara2}.
In spite of the small $M_h$, this set of SUSY parameters 
 is compatible with CERN LEP data in 
searching for neutral Higgs bosons of the MSSM~\cite{lephiggs},
and yields $\sin(\beta -\alpha)=0.33$.

\subsubsection{Signal and background Processes}

In this case, the signal event is
produced via
\begin{eqnarray}
  q + q^{\prime} & \rightarrow & A ( \rightarrow b \ov{b} )
  H^+ ( \rightarrow
  \nu
  \tau^+ ( \rightarrow \pi^+ {\bar \nu} )),\nonumber \\
  q + q^{\prime} & \rightarrow & H ( \rightarrow b \ov{b} )
  H^+ ( \rightarrow \nu
  \tau^+ ( \rightarrow \pi^+ {\bar \nu}  ))\, .
  \label{eq:signaltaunu}
\end{eqnarray}
The experimental signature of the signal events is the production of
$\pi^+ b {\bar b}$ associated with a large missing transverse momentum 
carried away by neutrino and anti-neutrino.
In this study we shall assume that both $b$ and ${\bar b}$ jets
in the signal event can be tagged with a total
efficiency of $50\%$, 
so that the four dominant SM background processes are:
\begin{eqnarray}
  W b \ov{b}: & & q q' \rightarrow b \ov{b} W^+, W^+
  \rightarrow \tau^+ \nu,
  \tau^+ \rightarrow \pi^+ {\bar \nu}  \, , \nonumber \\
  t \ov{b}:  & & q q' \rightarrow W^{\ast} \rightarrow t \ov{b},
    t \rightarrow b W^+, W^+
  \rightarrow \tau^+ \nu, \tau^+ \rightarrow \pi^+ {\bar \nu} \, ,
  \nonumber \\
  W g:  & & q g \rightarrow q' t {\bar b}, t \rightarrow W^+ b,
  W^+ \rightarrow
  \tau^+ \nu, \tau^+ \rightarrow \pi^+ {\bar \nu} \, ,
  \nonumber \\
  t \ov{t}:  & &  q q, \hspace{0.5em} g g \rightarrow t \ov{t}, t
  \rightarrow W^+ b, W^+ \rightarrow \tau^+ \nu, \tau^+ \rightarrow \pi^+
  {\bar \nu}, \ov{t} \rightarrow \ov{b} \hspace{0.1cm} W^- \, .
\label{eq:smbckg}
\end{eqnarray}
The first two processes are the intrinsic background processes
generated by the SM, and yield the same experimental signature as
the signal process at the parton level. In addition to the $W b
\ov{b}$ and $t \ov{b}$ processes, the other two important SM
background processes are $Wg$ and $t {\bar t}$. When the forward
jet ($q'$) in the $Wg$ fusion process is not detected (either
falling into a large rapidity region or carrying a too small
transverse momentum to be detected), it will mimic the
experimental signature of the signal event. Similarly, when the
decay products of $W^-$ in $\bar t \ra {\bar b} W^-$  escape
detection, the $t {\bar t}$ event yields the same event signature
as the signal. To study how often this situation happens, we
separately consider the leptonic and hadronic decay modes of $W^-$
at the parton level. It is straightforward to study the decay
modes $W^- \ra \ell^- {\bar \nu}_\ell$ with $\ell^- =e^-$ and
$\mu^-$. For $W^- \ra \tau^- {\bar \nu}_\tau$, we consider the
decay of $\tau^-$ into the $\ell^- {\bar \nu}_\ell {\bar
\nu}_\tau$, $\pi^- {\bar \nu}_\tau$ or $\rho^- {\bar \nu}_\tau$
mode. We find that the dominant contribution after imposing the
veto cuts (to veto events with extra charged lepton with large
transverse momentum or hadronic activities in the central rapidity
region) comes from $W^- \ra \ell^- {\bar \nu}_\ell$ and $W^- \ra
{\bar \nu}_\tau \tau^- (\ra \ell^- {\bar \nu}_\ell {\bar
\nu}_\tau)$,
 with $\ell^- =e^-$ and $\mu^-$.
Since in the end of analysis, the $t {\bar t}$ rate is smaller
than the $W b {\bar b}$ rate by more than one order of magnitude,
we shall defer the detailed discussion on this
part of the analysis to Sec.~III.C, where Case C is considered.
As to be shown later, for Case C, the dominant background comes from
the $t {\bar t} b {\bar b}$ events which mimic the
signal event signature when the decay products of
the extra $W^-$, from ${\bar t}$ decay, escape detection.

\subsubsection{Analysis}

\paragraph{Basic cuts:}
%\subsubsection{Basic cuts}

To compare the relevant background event rates to the signal event
rate, we shall assume the integrated luminosity of the LHC to be
$100\,{\rm fb}^{-1}$, and require the transverse momentum ($p_T^q$) and
the rapidity ($\eta^q$) of $b$, ${\bar b}$ and $\pi^+$ to satisfy
the following basic cuts: \beq
 p_T^q > 15 \,{\rm GeV}, \qquad |\eta^q| < 3.5, \qquad  \Delta R > 0.4  \,.
\label{eq:basic} \eeq Here, we have assumed a perfect detector
that can precisely measure the four-momenta of the final state
partons and require the separation in $\Delta R$ ($ \equiv
\sqrt{(\Delta \phi)^2 + (\Delta \eta)^2}$) between any two
observable final state partons (not including neutrinos) to be
larger than 0.4, where $\Delta \phi$ and $\Delta \eta$ are the
separation in azimuthal angle and rapidity, respectively. (We
shall comment on the effect due to the finite resolution of the detector
in the later part of this section.) Furthermore, in the $Wg$
event, the additional $q'$ jet (preferably in the forward
direction) is required to escape detection, i.e., either its
transverse momentum is less than 10\,GeV or its rapidity (in
magnitude) is larger than 3.5. Similarly, the charged lepton
$\ell^-$ ($= e^-, \mu^-, \tau^-$) from the decay of ${\bar t}$ in
the $t {\bar t}$ event is required to be undetected, i.e.,
$p_T^{\ell^-} < 10\,{\rm GeV}$ or $|\eta^{\ell^-}| > 3.0$. For
clarity, we shall only include the positively charged state (i.e.,
$\pi^+$) in the following discussion. The inclusion of the
opposite charged state (i.e., $\pi^-$) and other decay modes of
$\tau^+$ will be discussed in Sec. III D. The numbers of
the signal and background events after imposing the above cuts are
summarized in the second column of Table~\ref{tbl:data1}, in which
the $b$-tagging efficiency ($50\%$, for tagging both $b$ and
${\bar b}$ jets) is included. The last three rows in 
Table~\ref{tbl:data1} show
the ratio of signal ($S$) to background ($B$) event
number, the statistical significance of the
signal, and the statistical uncertainty in the measured signal
event rate.

\begin{figure}[t]
\begin{center}
\scalebox{0.6}{\includegraphics{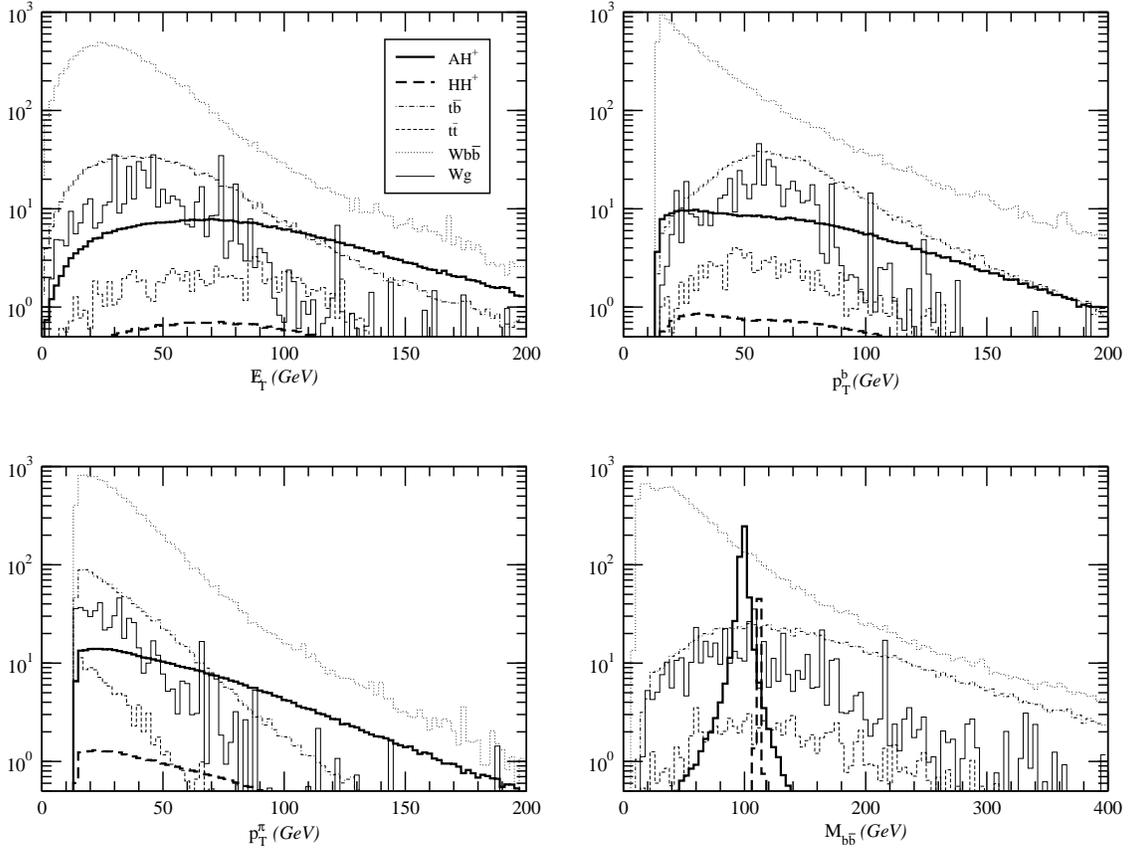}}
\end{center}
\caption{Distributions of
$\slash{E}_T$, $p_T^b$, $p_T^{\pi}$ and $m_{b\ov{b}}$ at the LHC
for Case A, with $M_A=101$\,GeV and $M_H=113$\,GeV
in the $b {\bar b} \pi^+ {\slash E_T}$ channel,
after imposing the basic cuts specified in Eq.~(\ref{eq:basic}).}
\label{fig:basic}
\end{figure}

At this stage of the analysis, the background rate is an order of
magnitude larger than the signal rate. Moreover, the dominant
background comes from the $Wb{\bar b}$ process, followed by the
$s$-channel single-top ($t {\bar b}$) process, the $t$-channel
single-top ($Wg$) process and the $t \bar t$ pair process. To
detect the signal event, further kinematic cuts are needed. In
order to study the efficient cuts that can largely suppress the
background rates while keeping most of the signal rates, we
examine the distributions of the missing transverse energy
($\slash{E}_T$), transverse momentum of $b$ jet ($p_T^b$),
transverse momentum of $\pi^+$ ($p_T^{\pi}$),
 and
the invariant mass of ${b\ov{b}}$ pair ($m_{b\ov{b}}$). Their
distributions are shown in Fig.~\ref{fig:basic}. 
(The vertical axis shows the
number of events per bin for all the figures presented in this
section.)

\begin{table}[t]
\caption{Numbers of $AH^+$ signal and background events for Case A,
with $M_A=101$\,GeV
in the $b {\bar b} \pi^+ {\slash E_T}$ channel, at the LHC with an
integrated luminosity of $100\,{\rm fb}^{-1}$. The $b$-tagging efficiency
($50\%$, for tagging both $b$ and ${\bar b}$ jets) is included, and
the kinematic cuts listed in each column are applied sequentially.}
\begin{tabular}{c|c|c|c|c|c}
\hline
\hline
 &Basic Cuts & $\slash E_T>50$GeV & $p_T^{\pi}>40$\,GeV &
 $90<M_{b\ov{b}}<110$\,GeV& With smearing\\

\hline
$ AH^+$    & 507   & 391   & 241   & 216  & 202   \\
\hline
$HH^+$    & 48    & 38    & 24    & 0    & 4     \\
$Wb\ov{b}$       & 11555 & 3111  & 864   & 67   & 62    \\
$t\ov{b}$        & 1228  & 614   & 163   & 12   & 11    \\
$Wg$             & 567   & 236   & 68    & 11   & 9     \\
$t\ov{t}$        & 110   & 80    & 17    & 2    & 2     \\
\hline
Signal ($S$)     & 507   & 391   & 241   & 216  & 202   \\
Bckg ($B$)       & 13507 & 4078  & 1135  & 92   & 87    \\
\hline
$S/B$            & 0.038 & 0.095 & 0.212 & 2.35 & 2.32  \\
$S/\sqrt{B}$     & 4.36  & 6.12  & 7.14  & 22.5 & 21.6  \\
$\sqrt{S+B}/S$   & 0.23  & 0.17  & 0.15  & 0.08 & 0.08  \\
\hline
\hline
\end{tabular}
\label{tbl:data1}
\end{table}

\begin{table}
\caption{Numbers of $HH^+$ signal and background events for Case A,
with $M_H=113$\,GeV
in the $b {\bar b} \pi^+ {\slash E_T}$ channel, at the LHC with an
integrated luminosity of $100\,{\rm fb}^{-1}$. The $b$-tagging efficiency
($50\%$, for tagging both $b$ and ${\bar b}$ jets) is included, and
the kinematic cuts listed in each column are applied sequentially.}
\begin{tabular}{c|c|c|c|c|c}
\hline
\hline
 &Basic Cuts & $\slash E_T>50$GeV & $p_T^{\pi}>40$\,GeV &
 $105<M_{b\ov{b}}<125$\,GeV& With smearing\\

\hline
$HH^+$    & 48    & 38    & 24    & 24   & 22 \\
\hline
$AH^+$    & 507   & 391   & 241   & 26   & 43   \\
$Wb\ov{b}$       & 11555 & 3111  & 864   & 58   & 54   \\
$t\ov{b}$        & 1228  & 614   & 163   & 11   & 11   \\
$Wg$             & 567   & 236   & 68    & 13   & 12   \\
$t\ov{t}$        & 110   & 80    & 17    & 2    & 2    \\
\hline
Signal ($S$)     & 48    & 38    & 24    & 24   & 22  \\
Bckg ($B$)       & 13966 & 4431  & 1352  & 110  & 120  \\
\hline
$S/B$            & 0.003 & 0.008 & 0.018 & 0.22 & 0.19 \\
$S/\sqrt{B}$     & 0.41  & 0.57  & 0.65  & 2.26 & 2.03 \\
$\sqrt{S+B}/S$   & 2.47  & 1.76  & 1.55  & 0.49 & 0.53 \\
\hline
\hline
\end{tabular}
\label{tbl:data2}
\end{table}

\paragraph{Missing $\slash{E}_T$:}
%\subsubsection{Missing $\slash E_T$}

As shown in Fig.~\ref{fig:basic}, the typical size of the
missing transverse energy in the signal event is larger than that in the
$Wb{\bar b}$, $t {\bar b}$ and $Wg$ background events. This is because
the main source of the missing energy in the signal and background
events comes from the neutrino in the decay of $H^+$ and $W^+$,
respectively, and
the mass of $H^+$ considered here is
larger than the mass of the $W^+$ boson.
The average value of $\slash E_T$ in the $t {\bar t}$ event is larger
than the other background processes, for the undetected
charged lepton $\ell^-$ from the decay of $\bar t$
also contributes to the missing energy of the event.
The same argument also explains why the average missing energy in the
$Wg$ event
(with an extra $q'$ parton jet)
is slightly larger than that in the $Wb {\bar b}$ and
$t {\bar b}$ events.
In Table~\ref{tbl:data1}, we show the number of signal and 
background events after
demanding the missing energy of the event to be larger than 50\,GeV.
This cut increases the signal-to-background ratio by about a factor
of 2.5 while keeping about $77\%$ of the signal rate.
The biggest reduction in the background rate comes from the
$Wb {\bar b}$ event (reduced by a factor of 3.7).

\begin{figure}
\begin{center}
{\scalebox{0.6}{\includegraphics{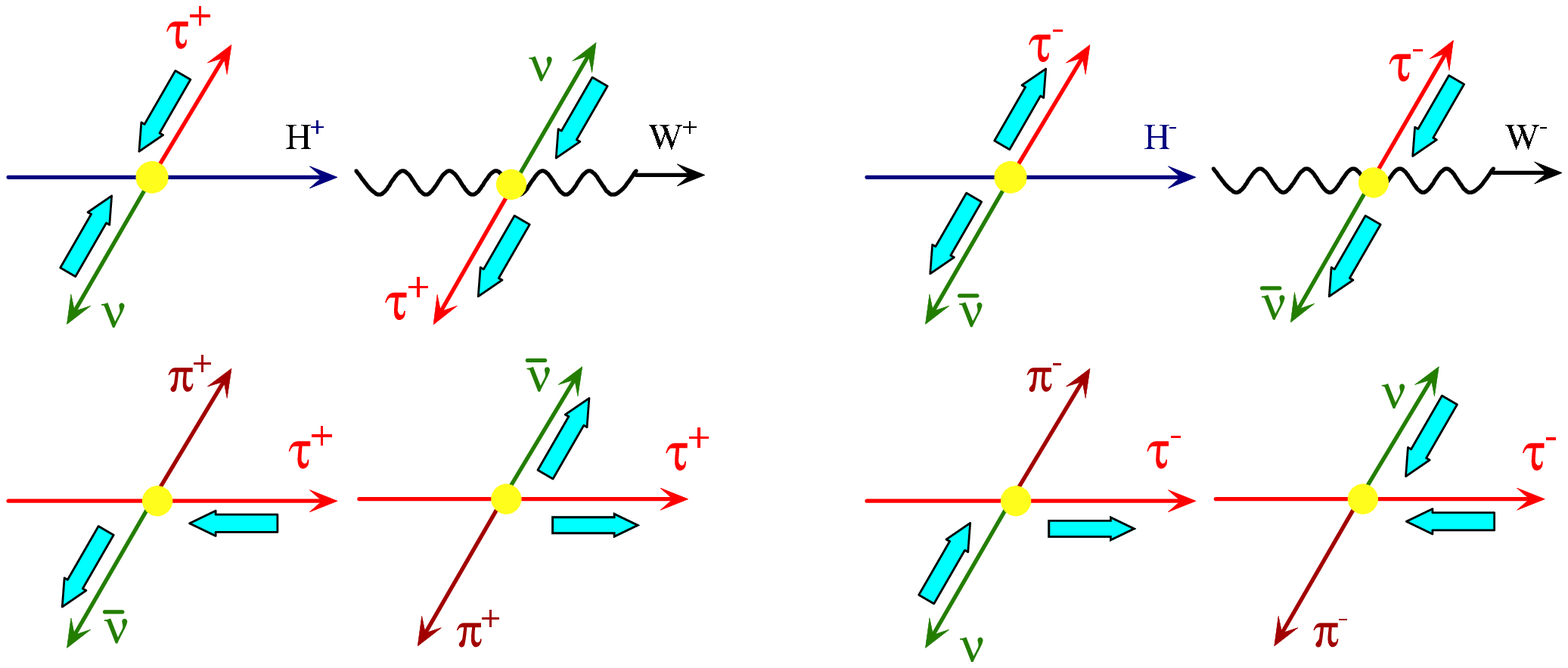}}}\\
\end{center}
\caption{$\tau^+$ is left-handedly polarized
in $H^+ \rightarrow \tau^+\nu$,
and right-handedly polarized in $W^+ \rightarrow \tau^+\nu$.
Moreover, $\pi^+$ momentum depends on the polarization of $\tau^+$.
(The thin arrow represents the moving direction, and 
the bold arrow represents the spin direction of the particle.)
Similar plots for $\tau^-$ are also shown in the right four diagrams.}
\label{fig:pol}
\end{figure}

\paragraph{Transverse momentum of the $\pi$-jet:}
%\subsubsection{Transverse momentum of $\pi$-jet}

Another big difference between the signal and the background
event signature is in the distribution of the transverse
momentum of $\pi^+$ originated from the decay of $\tau^+$, cf.
Fig.~\ref{fig:basic}. It can be understood as follows. In the
signal event, the coupling of $H^- \tau^+ \nu_\tau$ is almost
left-handed due to the enhancement from large $\tan \beta$.
Consequently, the $\tau^+$ from $H^+$ decay is produced
preferentially left-handedly polarized, cf. Fig.~\ref{fig:pol}. On
the contrary, in the background events, $\tau^+$ comes from the
decay of the weak gauge boson $W^+$ via left-handed weak current.
In the dominant background ($Wb {\bar b}$) event, the $W^+$ boson
is predominantly left-handed polarized. Therefore, $\tau^+$, as an
anti-fermion, is right-handedly polarized when ignoring the small
mass of $\tau^+$. In the weak decay of $\tau^+ \rightarrow \pi^+
{\bar \nu}_\tau$, the left-handed weak current forces ${\bar
\nu}_\tau$ to be purely right-handed in the limit that the
neutrino mass is ignored. Hence, to preserve angular momentum
conservation, $\pi^+$ would prefer moving along the direction of a
left-handed $\tau^+$, and against the direction of a right-handed
$\tau^+$, cf. Fig.~\ref{fig:pol}. Namely, the angular distribution
of $\pi^+$ is $(1+\cos\theta_{\pi})$ in the rest frame of a
left-handed $\tau^+$, and $(1-\cos\theta_{\pi})$ for a
right-handed, where $\theta_{\pi}$ is the polar angle of $\pi^+$
momentum with respect to the moving direction of $\tau^+$ in the
rest frame of $H^+$ (or $W^+$). Hence, the transverse momentum of
$\pi^+$ in $H^+ \rightarrow \nu_\tau \tau^+ (\rightarrow \pi^+
{\bar \nu}_\tau) $ is typically larger than that in $W^+
\rightarrow \nu_\tau \tau^+ (\rightarrow \pi^+ {\bar \nu}_\tau) $.
To further suppress the background rate, we require
$p_T^{\protect\pi} > 40$\,GeV. This cut increases the
signal-to-background ratio by a factor of 2.2 at the cost of the
reduction in signal rate by $40\%$, while the background rate is
reduced by a factor of 3.6. As clearly indicated in
Table~\ref{tbl:data1}, a similar reduction factor applies to all
different background processes.

\paragraph{Invariant mass of the $b {\bar b}$-jet pair:}
%\subsubsection{Invariant mass of $b {\bar b}$-jet pair}

\begin{figure}[t]
\begin{center}
\scalebox{0.4}{\includegraphics{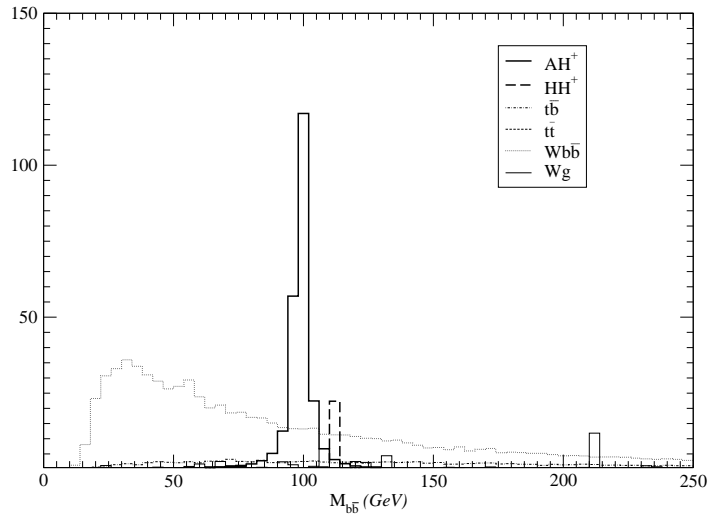}}
\end{center}
\caption{Distributions of $m_{b\ov{b}}$ for Case A,
with $M_A=101$\,GeV and $M_H=113$\,GeV
in the $b {\bar b} \pi^+ {\slash E_T}$ channel,
after imposing the additional cuts:~{}
$\slash E_T > 50$\,GeV and $p_T^{\protect\pi} > 40$\,GeV.}
\label{fig:bbcut2}
\end{figure}

At this stage of the analysis, the dominant background rate still comes
from the $Wb {\bar b}$ process whose rate is about 3.6 times
the $AH^+$ signal rate.
Since in the $Wb {\bar b}$ process, the  $b {\bar b}$ pair
originates from a virtual gluon conversion, the invariant mass of
$b {\bar b}$ pair ($m_{b \ov{b}}$) is generally small.
This is clearly illustrated in Fig.~\ref{fig:basic}.
A similar, though less dramatic,  result after imposing the
kinematic cuts
$\slash E_T > 50$\,GeV and
$p_T^{\protect\pi} > 40$\,GeV,
is also shown in Fig.~\ref{fig:bbcut2} for comparison.
To further improve the signal-to-background ratio, we
require the invariant mass of
$b {\bar b}$ pair to be within $M_{A} \pm  \sigma$, where
$\sigma = 10$\,GeV is
the expected experimental resolution for a 100 GeV Higgs boson
decaying into a
$b {\bar b}$ pair \cite{bbmass}.
As shown in Table~\ref{tbl:data1}, the $Wb {\bar b}$ background rate is reduced
by a factor of 13, and the other background rates are also reduced by an
order of magnitude. This yields $S/B=2.35$ and $S/\sqrt{B}=22.5$, with a
total of 216 signal events,
$
  q + q^{\prime} \rightarrow A ( \rightarrow b \ov{b} )
  H^+ ( \rightarrow
  \nu
  \tau^+ ( \rightarrow \pi^+ {\bar \nu} )) \, ,
$
produced at the LHC with an integrated luminosity of
$100\,{\rm fb}^{-1}$.

Up to now, we have focused our discussion on the $AH^+$
signal channel.
A similar analysis can also be performed for the $HH^+$
signal channel, whose result is shown in Table~\ref{tbl:data2}, where we
have required the invariant mass of  $b \bar b$ pair
to be within $M_{H} \pm  \sigma$
with $\sigma =10$\,GeV.
As shown in Table~\ref{tbl:susypara2}, since the coupling of
$W^\pm H H^\mp$ is suppressed by a factor of $\sin(\beta-\alpha)$
as compared to that of $W^\pm A H^\mp$ and $M_H > M_A$,
the production rate of $HH^+$ is expected to be
smaller than the $AH^+$ rate
by more than a factor of $\sin^2(\beta-\alpha)$.
For $M_A=101$\,GeV and $\tan\beta=40$, this reduction rate is
about 11.
Consequently, this $HH^+$ signal is difficult to detect, for the
background rate is still
larger by a factor of 5 even after
requiring $\slash E_T > 50$\,GeV and
$p_T^{\protect\pi} > 40$\,GeV,
cf. Table~\ref{tbl:data2} and Fig.~\ref{fig:bbcut2}.

We therefore conclude that for a small $M_A$, the mass of $A$ can be
determined from the $M_{b \bar b}$ distribution,
as shown in Fig.~\ref{fig:bbcut2}.
However, to measure the mass of $H$ via this channel is challenging.

\paragraph{Transverse mass of the charged Higgs Boson:}
%\subsubsection{Transverse mass of the charged Higgs Boson}

\begin{figure}[t]
\begin{center}
\scalebox{0.4}{\includegraphics{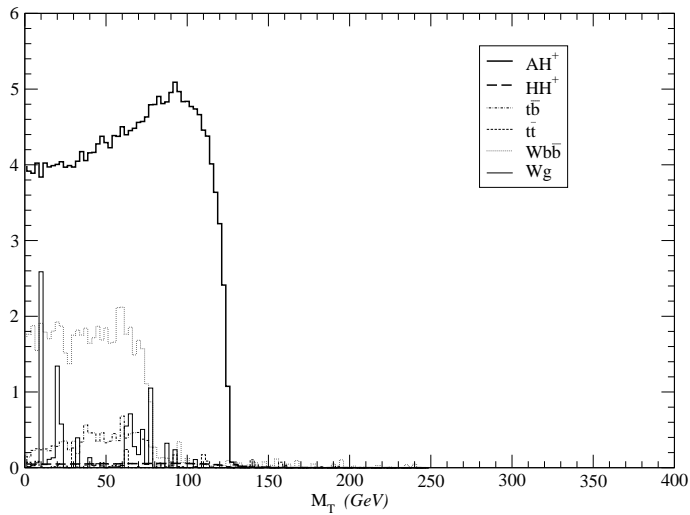}}
\end{center}
\caption{Distributions of the transverse mass $m_T$, defined in
Eq.~(\ref{eq:tm}), for Case A, after imposing the additional cut:
$|m_{b\ov{b}}-m_{A/H}|<10$\,GeV.} \label{fig:tmcut3}
\end{figure}

As discussed above, the mass of $A$ can be determined from the invariant
mass distribution of the $b \bar b$ pair. In order to test
the mass relation, Eq.~(\ref{eq:massrel}),
predicted by the MSSM, the mass of $H^+$ should also be measured.
In the signal event, the decay product of the charged Higgs boson $H^+$
contains the charged pion $\pi^+$ and the
 two (anti-)neutrinos ($\nu_\tau$ and ${\bar \nu}_\tau$), which
contribute to the missing transverse energy $\slash E_T$.
Hence, we can examine the transverse mass of $H^+$ defined as
\beq
 m_T = \sqrt{2 p_T^{\pi} \slash{E}_T ( 1 - \cos\phi )}
  \, ,\label{eq:tm}
\eeq
where $\phi$ is the azimuthal angle between the $\pi^+$-jet
and the missing transverse energy $\slash{E}_T$.
It is expected that
the shape of the $m_T$ distribution should exhibit a
Jacobian peak behavior when the
transverse momentum of $\pi^+$ is large, for
the dominant missing energy comes
from
$\nu_\tau$ in
$H^+ \ra \nu_\tau \tau^+( \ra \pi^+ {\bar \nu}_\tau) $.
This feature is clearly illustrated in Fig.~\ref{fig:tmcut3}.
The signal event shows a Jacobian peak around
$M_{H^+}$ (126\,GeV), while the dominant $W b {\bar b}$
background event shows a Jacobian peak around $m_W$ (80.2\,GeV)
but with a long tail into the larger $M_{b \bar b}$ region.
The long tail comes from the $W b {\bar b}$ events 
in which the transverse momentum of the $W^+$ boson is large.
Since the other background event rates are much smaller than
the $AH^+$ signal rate, we conclude that
the charged Higgs boson mass can be accurately extracted from the
transverse mass distribution.

\paragraph{Detector Effects:}
%\subsubsection{Detector Effects}

\begin{figure}[t]
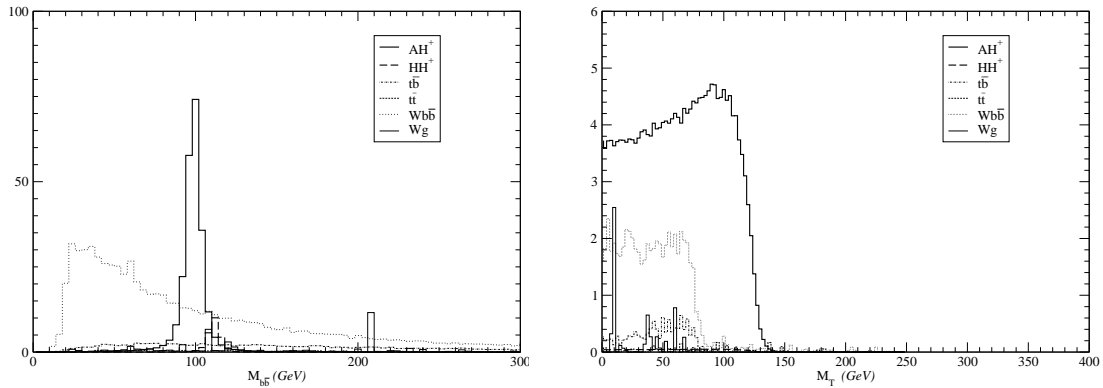

\begin{center}
\scalebox{0.3}{\includegraphics{bb_cut2_smearing.eps}}\qquad
\scalebox{0.3}{\includegraphics{tm_cut3_smearing.eps}}

\caption{Distributions of
invariant mass $m_{b\ov{b}}$ (after imposing the basic cuts,
$\slash E_T$ and $p_T^\pi$ cuts)
and transverse mass $m_T$ (after imposing all the kinematic cuts)
for Case A, with $M_A=101$\,GeV and $M_H=113$\,GeV,
after smearing the observable parton momenta by $50 \% \sqrt{E}$.}
\label{fig:smear}
\end{center}
\end{figure}

In reality, the performance of the detector is not perfect.
To study the effect due to the finite detection efficiency
of the detector, we repeat the above Monte Carlo analysis
after smearing all the final state parton momenta by
 a Gaussian distribution with
$$ \frac{\Delta E}{E} = \frac{50 \%}{\sqrt{E}} \, , $$
where $E$ is the energy of the observed parton and the resolution of the
energy measurement is assumed to be $50\% \sqrt{E}$.
The distributions of invariant mass $m_{b\ov{b}}$ and the
transverse mass $m_T$ become slightly broader, as shown in
Fig.~\ref{fig:smear}. However, both the signal and background rates,
as shown in the last column of Tables~\ref{tbl:data1} and \ref{tbl:data2},
are almost the same as those obtained with a perfect detector.

\begin{figure}[]
\begin{center}
\scalebox{0.3}{\includegraphics{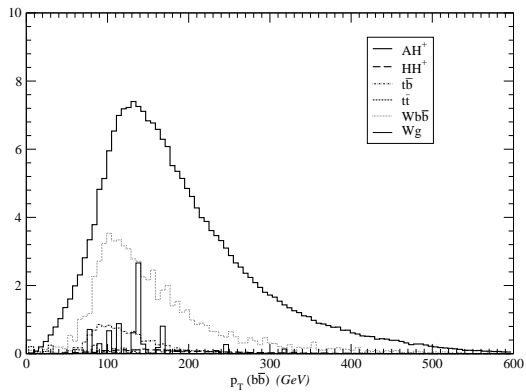}}
\caption{Transverse momentum distribution of the ${b\ov{b}}$ pair
for Case A, after imposing the additional cut:
$|m_{b\ov{b}}-m_{A/H}|<10$\,GeV.} \label{fig:ptbb}
\end{center}
\end{figure}

Finally, we note that in the above analysis we have limited ourselves to
a very minimal set of kinematic cuts to enhance the signal-to-background
ratio (by about a factor of 60) while keeping most of the signal
events (about half of them). There are other distributions that
can further distinguish signal from background events. For example,
as shown in Fig.~\ref{fig:ptbb},
 the transverse momentum of the $b \bar b$ pair in the $AH^+$
 signal events
 is typically larger than that in the background events, even after
 imposing the kinematic cuts $\slash E_T > 50$\,GeV and
$p_T^{\protect\pi} > 40$\,GeV.
This is because the scattering amplitude of the $AH^+$ signal event
is a p-wave amplitude, cf. Eq.~(3), and $M_{H^+}$ is larger than
$M_A$, so that $A$ is boosted to produce a large transverse momentum
which is typically about 100\,GeV in this case.


\subsection{$M_A=166$\,GeV}

\begin{figure}
\begin{center}
  \scalebox{0.3}{\includegraphics{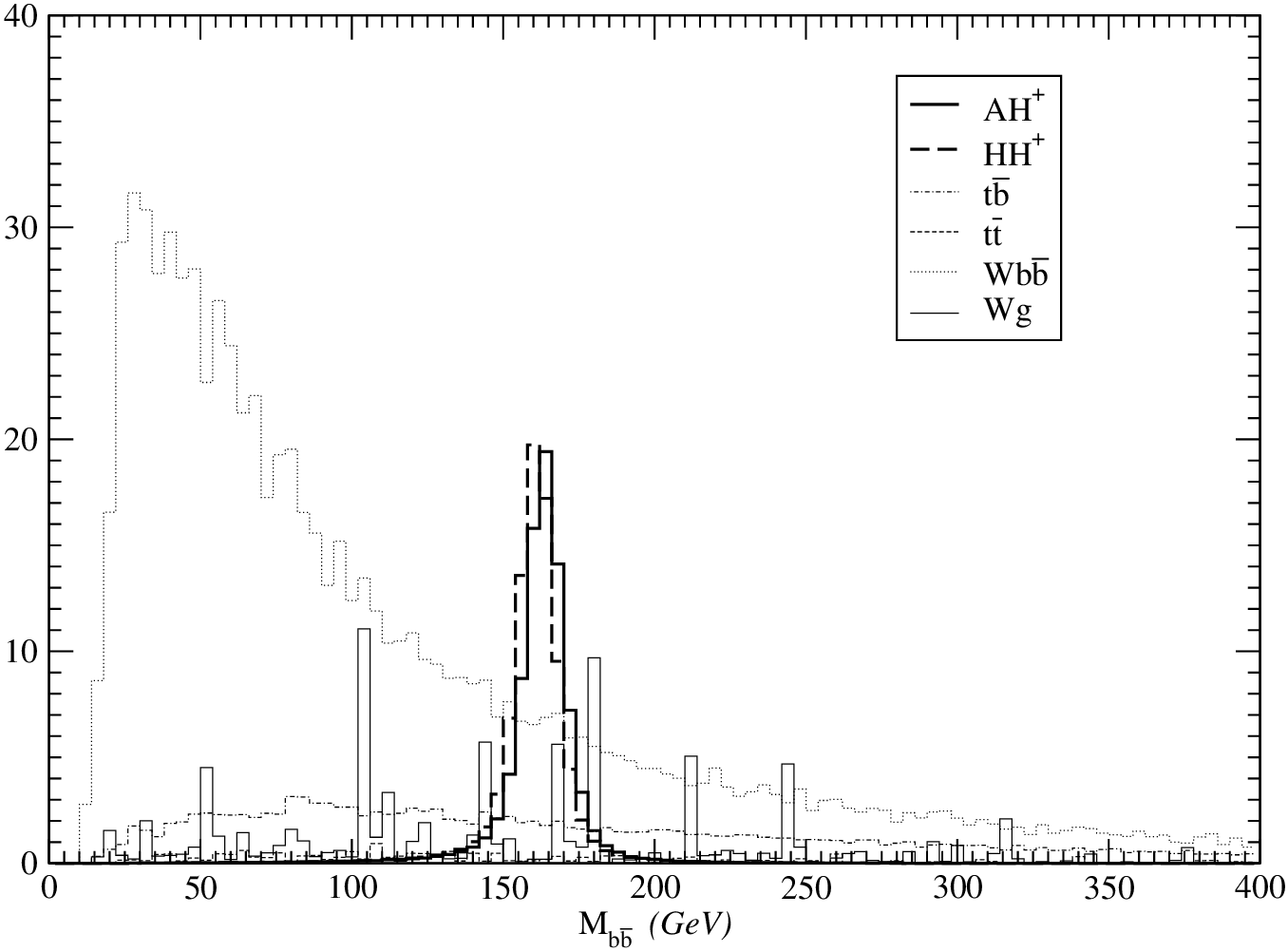}}\qquad
  \scalebox{0.3}{\includegraphics{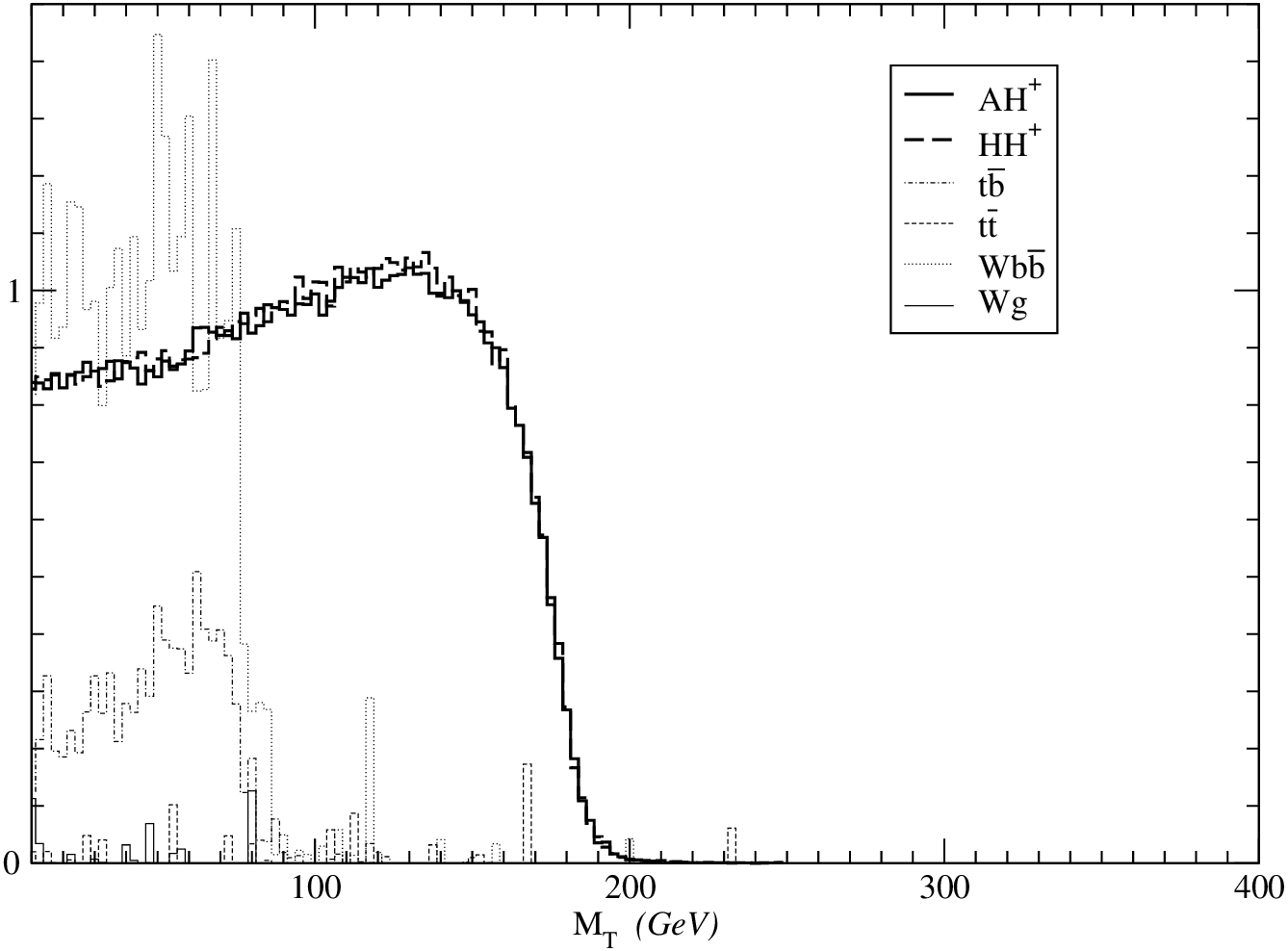}}
\end{center}
\caption{Distributions of
invariant mass $m_{b\ov{b}}$ (after imposing the basic cuts,
$\slash E_T$ and $p_T^\pi$ cuts)
and transverse mass $m_T$ (after imposing all the kinematic cuts) for
Case B, with $M_A=166$\,GeV and $M_H=163$\,GeV,
after smearing the observable parton momenta by $50 \% \sqrt{E}$.}
\label{fig:lsmear}
\end{figure}

We shall extend our analysis to the production of a charged Higgs
boson with its mass only slightly more than the top quark mass (174\,GeV)
so that the dominant decay
mode of $H^+$ remains to be the $\tau^+ \nu$ channel.
We take $M_A = 166$\,GeV, $\tan \beta = 40$, and all the other SUSY
parameters (soft-breaking masses and $\mu$ parameter)
to be 500\,GeV. Consequently,
the masses of the other Higgs bosons
are calculated from HDECAY~\cite{hdecay}, which gives
$M_h = 112 \textrm{~GeV}$,  
$M_H = 163 \textrm{~GeV}$ and $M_{H^+}
= 182 \textrm{~GeV}$. The relevant decay branching
ratios are $B ( A
\rightarrow b \ov{b} ) = 0.90$,
$B ( H \rightarrow b \ov{b} ) = 0 . 90$, $B ( H^+
\rightarrow \tau^+ \nu ) = 0 . 90$, and
$B ( \tau^+ \rightarrow \pi^+ \nu ) =
0.11$. The relevant total decay widths are $\Gamma_A = 5 .
6$\,GeV, $\Gamma_H = 5.5$\,GeV and $\Gamma_{H^{\pm}} = 0.68$\,GeV,
respectively. They are summarized in
Table~\ref{tbl:susypara2} for a quick reference.
In this case, $M_{H}$ is about the same as $M_A$, and
$\sin(\beta-\alpha) \sim 1$, hence, the production rates of
$AH^+$ and $HH^+$ are about the same.
Since  the decay branching ratios of
$A \ra b \bar b$
and
$H \ra b \bar b$
are also similar, we shall include both the
$AH^+$ and $HH^+$ event rates when we compute the signal rate
in this case.
Following the same analysis procedure as that for
the small $M_A$ case,
we show the numbers of signal and background events
in Table~\ref{tbl:data3}.
After imposing all the kinematic cuts, the
signal-to-background ratio increases by two orders of magnitude
 at the cost of half of signal events, and there are about
 140 signal events and 50 background events.
 In Fig.~\ref{fig:lsmear}, we show the distributions of the
invariant mass $m_{b\ov{b}}$ (after imposing the basic cuts,
$\slash E_T$ and $p_T^\pi$ cuts)
and transverse mass $m_T$ (after imposing all the kinematic cuts),
which are obtained after smearing the observable parton
momenta by $50 \% \sqrt{E}$ to mimic the effect of the finite 
resolution of the detector.

\begin{table}[t]
\caption{Numbers of $AH^{+}$ and $HH^{+}$ signal and background events for
Case B, with $M_A=166$\,GeV and $M_H=163$\,GeV
in the $b {\bar b} \pi^+ {\slash E_T}$ channel, at the LHC with an
integrated luminosity of $100\,{\rm fb}^{-1}$. The $b$-tagging efficiency
($50\%$, for tagging both $b$ and ${\bar b}$ jets) is included, and
the kinematic cuts listed in each column are applied sequentially.}
  {\begin{tabular}{c|c|c|c|c|c}
\hline
     & Basic Cuts & $\slash{E}_T>50$\,GeV&$p_T^{\pi}>40$\,GeV&
     $155<M_{b\ov{b}}<175$\,GeV& With smearing\\
\hline
    $A H^+$        & 126   & 111   & 85   & 71  & 65  \\
    $H H^+$        & 129   & 114   & 87   & 72  & 66  \\
\hline
    $Wb\ov{b}$       & 11560 & 3102  & 840  & 33  & 34  \\
    $t\ov{b}$        & 1221  & 607   & 164  & 10  & 10  \\
    $Wg$             & 783   & 318   & 11   & 5   & 6   \\
    $t\ov{t}$        & 108   & 79    & 18   & 1   & 1   \\
\hline
    Signal ($S$)          & 255   & 225   & 171  & 143 & 131 \\
    Bckg ($B$)             & 13672 & 4106  & 1031 & 48  & 50  \\
\hline
    $S/B$            & 0.02  & 0.05  & 0.17 & 3.0 & 2.6 \\
    $S/\sqrt{B}$     & 2.18  & 3.51  & 5.33 & 20.8& 18.6\\
    $\sqrt{S + B}/S$ & 0.46  & 0.29  & 0.20 & 0.09& 0.10\\
\hline
  \end{tabular}}
\label{tbl:data3}
\end{table}

\subsection{$M_A=250$\,GeV}

When the mass of the $H^+$ is larger than
 the sum of the masses of top and bottom
quarks but not too large that the supersymmetric decay modes
become significant, the dominant decay mode of $H^+$ is likely to be
$H^+ \rightarrow t \ov{b}$. Here, we study how to detect the
signal event in this decay mode.

To simplify the discussion, we shall concentrate on the semi-leptonic
decay mode of top quark, and the $b \bar b$ decay mode of $A$ or $H$.
Hence, the signal events considered here are produced via
\begin{eqnarray}
  q q^{\prime} & \rightarrow & A ( \rightarrow b \ov{b} ) H^+ ( \rightarrow
  \ov{b}  t ( \rightarrow b\hspace*{0.1cm} \ell^+\hspace*{0.1cm} \nu )),
  \nonumber \\
  q q^{\prime} & \rightarrow & H ( \rightarrow b \ov{b} ) H^+ ( \rightarrow
  \ov{b}  t ( \rightarrow b\hspace*{0.1cm} \ell^+\hspace*{0.1cm} \nu
  ))\, ,
 \label{eq:signaltb}
\end{eqnarray}
where $\ell^+=e^+$ or $\mu^+$.
The signature of the signal event is four $b$-jets plus one isolated
lepton and missing transverse energy.
To detect the signal event, we require the four $b$-jets
and the isolated charged lepton to be well
separated in $\Delta R$ with large $p_T$ in the central rapidity
region, in addition to a $\slash E_T$ signature.
In the following, we assume that the
efficiency for tagging all four $b$-jets is
$25\%$, which is included in the
calculation of the event rates to be discussed below.
The dominant SM background for such kind of signature comes from the
 $t\bar{t} b \bar{b}$ events, produced via
\begin{eqnarray}
  q \ov{q}, g g\rightarrow b \ov{b} \hspace*{0.1cm}
  t ( \rightarrow b \hspace*{0.1cm} \ell^+\hspace*{0.1cm} \nu )
  \ov{t}(\rightarrow \ov{b}\hspace*{0.1cm} W^-).
\label{eq:bkgdttbb}
\end{eqnarray}
When the decay products of $W^-$ escape detection, this background event
will mimic the signal event signature.
At the LHC, the $t\bar{t} b \bar{b}$ rate is dominated by the gluon
fusion process due to its large parton luminosity.
As shown in Table~\ref{tbl:tb}, after imposing the following basic kinematic
cuts:\footnote{
Here, we impose a set of stronger cuts than those given in 
Eq.~(\ref{eq:basic}) in order to have a more reliable estimate of the 
$t\bar{t} b \bar{b}$ background rate.
}
\begin{eqnarray}
p_T^{\ell^+}&>&15  \,{\rm GeV}, \quad |\eta^{\ell^+}|<3.0, \quad
{\slash E_T}> 20  \,{\rm GeV}\,,
\nonumber\\
p_T^{b}&>&20 \,{\rm GeV}, \quad |\eta^{b}|\,\,\,<2.0 \, ,\nonumber \\
\Delta R &>& 0.4 \, ,
\label{eq:basiccuttb}
\end{eqnarray}
the $gg \ra t\bar{t} b \bar{b}$ rate is larger than the
$q \bar q \ra t\bar{t} b \bar{b}$ rate by two orders of
magnitude.
Here, we have required the separation in $\Delta R$
between any two final state partons
among the four $b$-jets and the charged lepton $\ell^+$
to be larger than 0.4.

The signal rate depends on SUSY parameters. We take
$M_A = 250$\,GeV, $\tan \beta = 40$, and all the other SUSY
parameters (soft-breaking masses and $\mu$ parameter)
to be 500\,GeV. Consequently,
$M_H = 248\textrm{~GeV}$ and $M_{H^+}
=261\textrm{~GeV}$. The relevant decay branching ratios are $B ( A
\rightarrow b \ov{b} ) = 0.89$,
$B ( H \rightarrow b \ov{b} ) = 0.89$, $B ( H^+
\rightarrow t \ov{b} ) = 0.79$, and
$B ( t \rightarrow b\hspace*{0.1cm} e^+ \nu ) =
0.11$. The relevant total decay widths are $\Gamma_A = 7.9$\,GeV,
$\Gamma_H = 7.8$\,GeV and $\Gamma_{H^{\pm}} = 4.2$\,GeV, as summarized in
Table~\ref{tbl:susypara2}.
For this set of parameters, the event rates of the $AH^+$ and $HH^+$
modes are about the same due to the decoupling limit, i.e., $M_A$ becomes
large. Furthermore, the total signal rate
(the sum of the $AH^+$ and $HH^+$ rate) is about one hundredth of the
 $t\bar{t} b \bar{b}$ background rate, cf. Table~\ref{tbl:tb}, after the
 basic cuts, and the sum of the $AH^+$ and $HH^+$ signal
rates yields about 36 events at the LHC. We have checked that our
calculation of the $t\bar{t} b \bar{b}$ cross section agrees with
that in Refs.~\cite{ttbb,topquark}, where we have chosen the scale for
evaluating parton distribution functions and strong coupling to be
the invariant mass of the final state particles. (The $t\bar{t} b
\bar{b}$ cross section can vary as much as $50\%$ with a different
choice of the scale~\cite{ttbb}.)

\begin{table}[t]
\caption{Numbers of $AH^{+}$ and $HH^{+}$ signal and background events for
Case C, with $M_A=250$\,GeV and $M_H=248$\,GeV
in the $b {\bar b} b {\bar b} \ell^{+} {\slash E_T}$ channel,
at the LHC with an
integrated luminosity of $100\,{\rm fb}^{-1}$. The $b$-tagging efficiency
($25\%$, for tagging all the four $b$ jets) is included, and
the kinematic cuts listed in each column are applied sequentially.}
\label{tbl:tb}
\begin{tabular}
[c]{c|c|c}
\hline
\hline
& Basic Cuts &Veto Cuts\\
\hline
$AH^{+}$ & 18 &18\\
$HH^{+}$ & 18 & 18\\
\hline
\hline
$gg\rightarrow t\bar{t}b\bar{b}$ & 1632 & 3\\
\hline
$qq\rightarrow t\bar{t}b\bar{b}$ & 105 & 0\\
\hline
\hline
    Signal ($S$)          &  36   & 36  \\
    Bckg ($B$)            & 1737  & 3   \\
\hline
\hline
\end{tabular}
\end{table}

As noted above, the $t\bar{t} b \bar{b}$ background events contain
an extra $W^-$ boson in the final state, as compared to the signal
event. To suppress this large background event, we need to veto the
additional $W^-$ boson which can decay into di-jets or a charged lepton
plus neutrino. When $W^-$ decays into the di-jet mode, there will be
additional hadronic activities in the central rapidity region, hence,
one needs to apply kinematic cuts to veto the additional
high $p_T$ jets in the central rapidity region.
When $W^-$ decays into a charged lepton plus neutrino, an
additional charged lepton is produced in the final state.
Below, we first discuss the case that $W^-$ decays into
the leptonic mode with the charged lepton
($\ell^-$) being $e^-$ or $\mu^-$,
 which accounts for two ninth of the top quark decay branching
ratio. We veto the additional charged lepton with
a transverse momentum exceeding 10\,GeV and rapidity (in
magnitude) less than 3.0.
It turns out that the veto efficiency is so high
(around $99\%$) that the
background rate is down by three orders of magnitude
and yields 2 events of
${\bar t} \ra {\bar b} W^- (\ra {\nu} \ell^-)$ with $\ell^-=e^-$ or
$\mu^-$.
Consequently, the signal
rate is about one order of magnitude larger than this
background rate.
Though the above analysis is adequate for $\ell^- = e^-$ or $\mu^-$,
it is less accurate for $\ell^- = \tau^-$ when $\tau^-$ decays further
to either leptons or hadrons.
(The decay branching ratio of $W^- \ra {\bar \nu} \tau^-$
is about one ninth.)
Hence, we improve the calculation of the
${\bar t} \ra {\bar b} W^- (\ra {\bar \nu} \tau^-)$
background rate by considering the
leptonic decay mode $\tau^- \ra \ell^- \nu_\tau {\bar \nu}$
(with $\ell^- = e^-$ or $\mu^-$)
and the hadronic decay modes of tau, such as
$\tau^- \ra \pi^- \nu_\tau$ or $\rho^- \nu_\tau$.
For the
$\tau^- \ra \ell^- \nu_\tau {\bar \nu}$ decay mode,
we veto events in which
$p_T^{\ell^-}>10$\,GeV and  $|\eta^{\ell^-}|<3.0$.
After taking into account the
decay branching ratio of tau into the leptonic mode
(about $35\%$), this amounts to 1 event of
${\bar t} \ra {\bar b} W^-
(\ra {\bar \nu} \tau^-(\ra \ell^- \nu_\tau {\bar \nu}))$
passing through the vetoing cut.
It is straightforward to find out the background rate of
${\bar t} \ra {\bar b} W^-
(\ra {\bar \nu} \tau^-(\ra \pi^- \nu_\tau))$,
after vetoing the additional central jet (i.e., $\pi^-$ jet)
that has transverse momentum larger than 10\,GeV and rapidity (in
magnitude) less than 3.5.
We find that the vetoing efficiency for the
 $\tau^-\ra \pi^- \nu_\tau$ event, whose
decay branching ratio is about $11\%$, is so high that
its contribution to the $t {\bar t} b {\bar b}$ background rate is
negligibly small. The similar conclusion also holds for
the $\tau^- \ra  \nu_\tau \rho^- (\ra \pi^- \pi^0)$ mode,
whose decay branching ratio is about $25\%$.
For this decay mode,
we first check whether the separation of $\pi^-$ and $\pi^0$
in $\Delta R$ is within 0.4. If yes,
we sum up their momenta to form one big jet
and follow the same analysis as that for
$\tau^- \ra   \pi^- \nu_\tau$.
If not,
we veto any
additional central jet (i.e., $\pi^-$ or $\pi^0$ jet)
that has transverse momentum larger than 10\,GeV and rapidity (in
magnitude) within 3.5.
In order for the
${\bar t} \ra {\bar b} W^- (\ra {\bar q} q')$
background event to mimic the signal event,
the decay products of
$W^-$ have to escape detection.
When the separation of ${\bar q}$ and $q'$
in $\Delta R$ is within 0.4,
we sum up their momenta to form one big jet
and veto the event when the
transverse momentum of the big jet is
larger than 10\,GeV and rapidity (in magnitude) smaller than 3.5.
Otherwise,
we veto any additional central jet (i.e., ${\bar q}$ or $q'$ jet)
that has transverse momentum larger than 10\,GeV and rapidity (in
magnitude) within 3.5. This amounts to a negligibly small number
of ${\bar t} \ra {\bar b} W^- (\ra {\bar q} q')$ events at the
LHC, for its vetoing efficiency is close to $100\%$. In
Table~\ref{tbl:tb}, we sum up all the decay modes of $W^-$
discussed above to estimate the background rate $t\bar{t} b
\bar{b} \ra b {\bar b} b {\bar b} \ell^{+} {\slash E_T}$ which is
about one order of magnitude smaller than the signal rate.
Furthermore, the distribution of the transverse mass of $\ell^+$
and $\slash E_T$, which is the transverse mass of the $W^+$ boson
in the signal event, can be used to further discriminate the
signal from the background events. This is because the vetoing
cuts generate additional missing energy in the background event,
while in the signal event, the $\slash E_T$ is entirely generated
by the neutrino from $W^+$ boson decay. It is desirable to check the
above estimate using a full event generator
 (including effects of parton showering and hadronization)
combined with a realistic detector simulation, which is
however beyond the scope of this paper.

\begin{figure}[t]
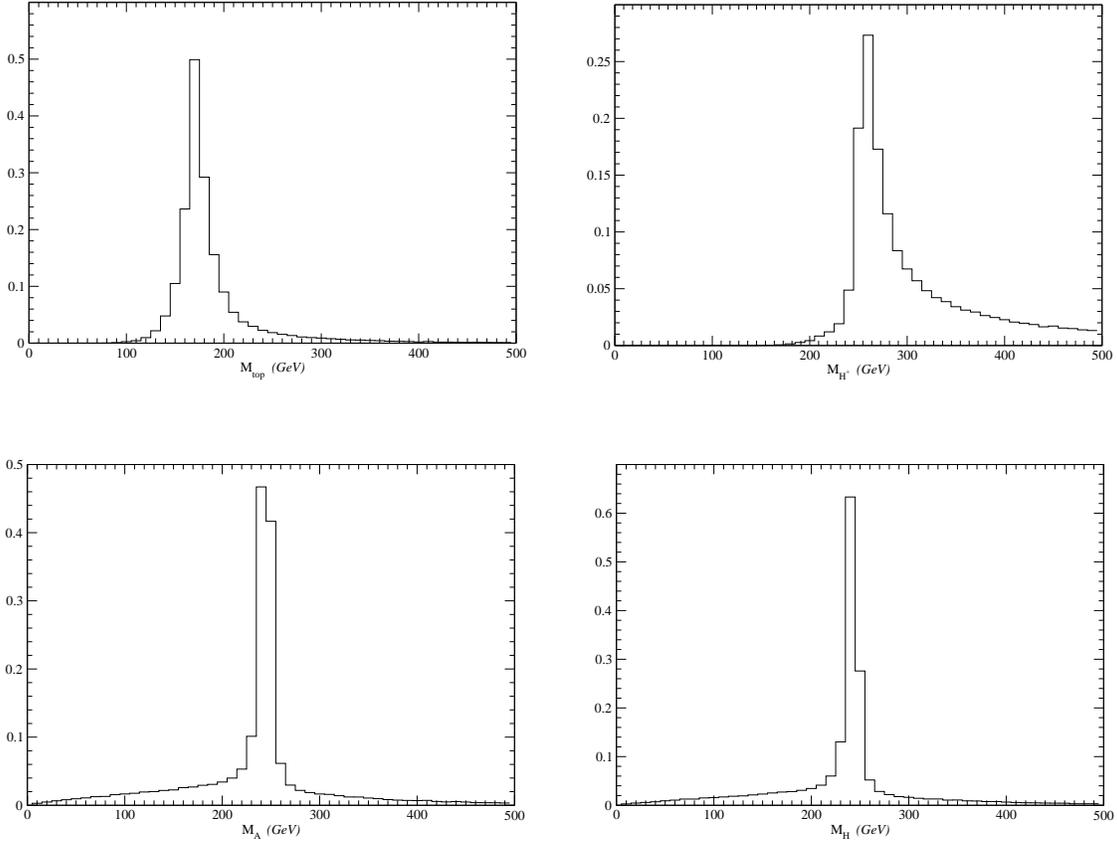

\begin{center}
\scalebox{0.3}{\includegraphics{wahtb_t.eps}}\qquad
\scalebox{0.3}{\includegraphics{wahtb_ch.eps}} \\
\vspace{1cm}
\scalebox{0.3}{\includegraphics{wahtb_a.eps}} \qquad
\scalebox{0.3}{\includegraphics{whhtb_h.eps}}
\caption{The invariant mass distributions of the reconstructed
(a) top quark $t$, (b) charged Higgs boson $H^+$,
(c) CP-odd Higgs Boson $A$, and (d) CP-even heavy Higgs boson $H$,
for Case C, with $M_{H^+}=261$\,GeV, $M_A=250$\,GeV and $M_H=248$\,GeV,
after
imposing all the kinematic cuts discussed in the text.}
\label{fig:tbnosmear}
\end{center}
\end{figure}

Next, we illustrate how to reconstruct the mass of $A$ and
$H^+$ in the signal event
$A ( \rightarrow b \ov{b} ) H^+ ( \rightarrow
  \ov{b}  t ( \rightarrow b\hspace*{0.1cm} \ell^+\hspace*{0.1cm} \nu ))
  \, .
$
A similar method also applies to the $HH^+$ signal event.
Since both $\ell^+$ and $\nu$ come from an on-shell $W^+$ boson decay,
we can use the $W$-boson mass constraint and
$\slash{E}_T$ information to specify the longitudinal momentum
($p_z(\nu)$) of the neutrino.
The above procedure leads to two possible solutions of $p_z(\nu)$.
Sometimes, only one of the two solutions is kinematically allowed, but 
most of the time both of them are physical solutions for a signal event. 
Therefore,
one has to fix a prescription~~\cite{Kane:1989vv}
 to choose the one which will most likely
give the correct distribution of the invariant mass of $e^+$, $\nu_e$ and $b$. 
Since $W$-boson, top quark and $H^{+}$ boson 
are all heavy particles, they 
and their decay products are likely to be produced in the central 
rapidity region. Hence, we choose the one with the
smaller magnitude in $p_z(\nu)$ to reconstruct the $W^+$ boson.
In the signal event, there are four $b$-jets.
(We consider the case that a ${\bar b}$-jet cannot be experimentally
distinguished from a ${b}$-jet.)
We pair the reconstructed
$W^+$ boson with any one of the four $b$-jets to calculate the invariant
mass of $W^+ b$ and choose the $b$-jet that yields a mass
closest to the nominal top mass (174\,GeV)
to be the one produced from the decay of
$t$. In Fig.~\ref{fig:tbnosmear}(a), we show the distribution of
the reconstructed
top quark mass which peaks around 174\,GeV with a
17\,GeV half-of-maximum width.
In order to pair the two correct $b$-jets, among the remaining three
$b$-jets, to reconstruct the mass of the CP-odd Higgs boson $A$, we make
use of the mass inequality implied by the MSSM mass relation,
Eq.~(\ref{eq:massrel}).
Namely, the mass of $A$ is smaller than the mass of $H^+$. (As $M_A$
becomes large, the mass of $H$ is about the same as the mass of $A$.)
We loop over the remaining three $b$-jets (labelled as $b_{1,2,3}$), and
calculate the invariant mass of $t b_1$, denoted as
$m_{t b_1}$, and the invariant mass of $b_2 b_3$, denoted as
$m_{b_2 b_3}$. We choose the assignment of the three $b$-jets so that
$m_{t b_1} > m_{b_2 b_3}$ and
the difference in $m^2_{t b_1}$ and $m^2_{b_2 b_3}$ is closest to
$M^2_W$.
Following this procedure, we obtain the distributions of $M_A$ and
$M_{H^+}$ which are shown in
Figs.~\ref{fig:tbnosmear}(b) and~\ref{fig:tbnosmear}(c).
The distribution of the reconstructed $M_A$
peaks around 250\,GeV with a 10\,GeV, half-of-maximum
width, and the reconstructed $M_{H^+}$
peaks around 261\,GeV with a 20\,GeV half-of-maximum
width.
The broad width in the reconstructed $M_{H^+}$ distribution is caused
by the wrong assignment of the $b$-jets, for the difference in $M_{H^+}$
and $M_A$ (about 13\,GeV) is at the same order as the decay widths of
$A$ and $H^+$.
Hence, even with the crude analysis discussed above,
 it is possible to reconstruct the mass of
$A$ and $H^+$ in the $AH^+ \ra b \ov{b} \ov{b}  t $ signal event.
Similarly, $M_H$ can also be reconstructed for the
$HH^+ \ra b \ov{b} \ov{b}  t $ signal event, whose result is shown in
Fig.~\ref{fig:tbnosmear}(d).
A more complicated analysis using the likelihood method,
such as the neural network analysis, would certainly improve the
above conclusion.

\begin{figure}
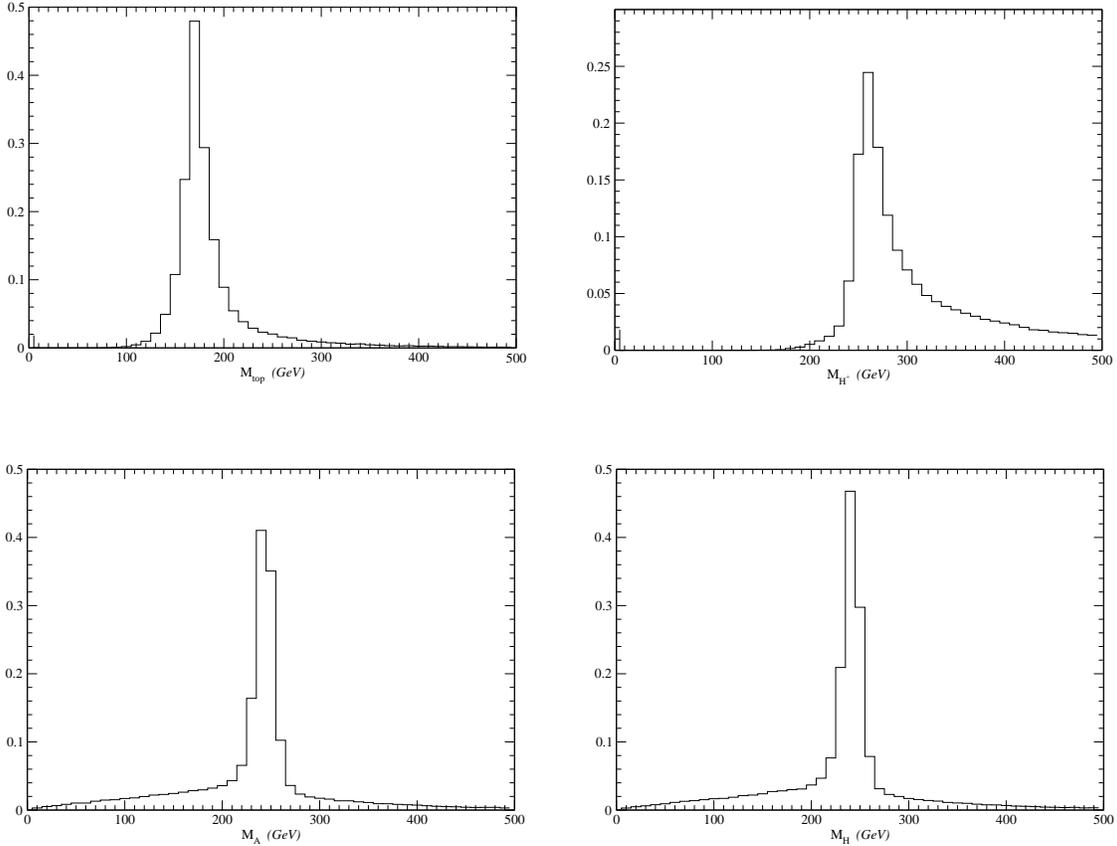

\begin{center}
\scalebox{0.3}{\includegraphics{wahtb_t_smear.eps}}\qquad
\scalebox{0.3}{\includegraphics{wahtb_ch_smear.eps}} \\
\vspace{1cm}
\scalebox{0.3}{\includegraphics{wahtb_a_smear.eps}} \qquad
\scalebox{0.3}{\includegraphics{whhtb_h_smear.eps}}
\caption{The invariant mass distributions of the reconstructed
(a) top quark $t$, (b) charged Higgs boson $H^+$,
(c) CP-odd Higgs Boson $A$, and (d) CP-even heavy Higgs boson $H$,
for Case C, with $M_{H^+}=261$\,GeV, $M_A=250$\,GeV and $M_H=248$\,GeV,
after smearing the observable parton momenta by $50 \% \sqrt{E}$ and
imposing all the kinematic cuts discussed in the text.}
\label{fig:tbsmear}
\end{center}
\end{figure}

Finally, we comment on the effect due to the finite resolution of the
detector. Again, we smear the four-momenta of the final state
partons (except neutrino) by
assuming a Gaussian error of $50\% \sqrt{E}$, where $E$ is the energy of the
parton. Repeating the above analysis, we
obtain the distributions of the reconstructed
$M_t$, $M_A$ (or $M_H$) and $M_{H^+}$
as shown in Fig.~\ref{fig:tbsmear}.
Except the small difference in the height of the peak, they 
are similar to those obtained for a perfect 
detector, as shown in Fig.~\ref{fig:tbnosmear}.

\subsection{Discussions}

Before concluding this section, we would like to
summarize the results of our Monte Carlo analysis
and discuss the need for a realistic simulation
of the detector and a full event generator analysis.
In this work, we considered three different signal scenarios
as given in Table~\ref{tbl:susypara2}.
In Case A and Case B, the dominant decay mode of $H^+$ is
$H^+ \ra \tau^+ \nu_\tau$. For Case A, the
SUSY parameters of the Higgs sector are
in the mixing regime
[$\sin(\beta-\alpha) \sim 0.3$-$0.8$], so that
$M_H$ differs from $M_A$.
For Case B, the SUSY parameters of the Higgs sector
are in the decoupling regime
[$\sin(\beta-\alpha) \sim 1$], so that
$M_H$ is almost the same as $M_A$.
In Case C, the $t \bar b$ mode opens and becomes the dominant decay
channel of $H^+$, so that it provides a different event signature.
The results of our Monte Carlo analysis are
separately summarized as follows.

To detect the signal events for Case A and Case B, we consider the
decay modes of $A \ra b \bar b$ (or $H \ra b \bar b$)
 and $H^+ \ra \tau^+ \nu_\tau$ with
$\tau^+ \ra \pi^+ {\bar \nu}_\tau$ in the $AH^+$ (or $HH^+$)
events.
The relevant decay widths and branching ratios are summarized in
Table~\ref{tbl:susypara2} for a quick reference.
The dominant SM background events for these cases come from
$W b \bar b$, $t \bar b$, $Wg$, and $t \bar t$ events,
as defined in Eq.~(\ref{eq:smbckg}).

Firstly, we discuss Case A, in which the production rate of $HH^+$
is smaller than $AH^+$ by a factor of $\sin^2(\beta - \alpha) \sim
0.11$. After imposing the basic cuts, specified in
Eq.~(\ref{eq:basic}), and assuming a $50\%$ of $b$-tagging
efficiency for tagging both $b$-jets in the event, we obtain the
signal and background event rates as shown in the second column of
Table~\ref{tbl:data1} for the LHC with an integrated luminosity of
$100\,{\rm fb}^{-1}$. By noting that the background events
(particularly the $W b \bar b$ events) tend to have smaller
missing energy, we impose a cut, $\slash E_T > 50\,{\rm GeV}$, to
suppress various background rates. The results are shown in the
third column of Table~III. Furthermore, the polarization of
$\tau^+$ in the signal events (left-handedly polarized) is
different from that in the SM background events (right-handedly
polarized). Consequently, the typical transverse momentum of
$\pi^+$ from the decay of $\tau^+$ is higher in the signal events
than in the SM background events, cf. Fig.~\ref{fig:pol}. Hence,
we impose another cut, $p_T^{\pi} > 40$\,GeV, to further suppress
the background rates and increase the signal-to-background ratio.
This result is shown in the fourth column of
Table~\ref{tbl:data1}. To detect the $AH^+$ signal event, we
select the invariant mass of the $b \bar b$ pair so that it is within
10\,GeV around $M_A$. The result is shown in the fifth column of
Table~\ref{tbl:data1}. As shown, after these cuts, the
significance of the signal event increases by a factor of 5 at the
cost of about half of the number of signal events. A similar
result for detecting the $HH^+$ signal event is given in
Table~\ref{tbl:data2}. Up to now, the analysis was carried out at
the parton level with a perfect measurement of the observable
parton momenta. In reality, the detection efficiency of the
detector cannot be one hundred percent. To estimate the effect due
to the finite detection efficiency of the detector, we smear
the four-momenta of the final state observable partons (excluding
neutrino) by assuming a Gaussian error of $50\% \sqrt{E}$, where
$E$ is the energy of the parton. After repeating the above
analysis with the basic cuts, the $\slash E_T$, $p_T^{\pi}$, and
$m_{b \bar b}$ cuts, we obtain the signal event rates as shown in
the last column of Table~\ref{tbl:data1} and Table~\ref{tbl:data2}
for the $AH^+$ and $HH^+$ channels, respectively. It is clear that
the $AH^+$ signal is much easier than the $HH^+$ signal to be
identified for Case A. Furthermore, from the distribution of the
$b \bar b$ invariant mass $m_{b \bar b}$, one can extract out the
$A$ boson mass $M_A$ from its peak position, cf.
Fig.~\ref{fig:bbcut2}. From the distribution of the transverse
mass of $H^+$ defined as Eq.~(\ref{eq:tm}), one can extract out
the $H^+$ boson mass $M_{H^+}$ from its jacobian peak position,
cf. Fig.~\ref{fig:tmcut3}. Given the measured $M_A$ and $M_{H^+}$
values, one can then test the mass relation,
Eq.~(\ref{eq:massrel}), to confirm the MSSM Higgs sector. This is
the conclusion for the case that a $AH^+$ signal, as predicted by
Case A, is found. If it is not found, then one can readily
constrain the MSSM. For example, according to the last column in
Table~\ref{tbl:data1}, if a $AH^+$ signal is not found, then at
the two standard deviation level, the product of $B(A \ra b \bar
b)$ and $B(H^+ \to \nu_\tau \tau^+)$ in the MSSM has to be less than
$\displaystyle \frac{2 \sqrt{87}}{202} \times (0.91)(0.98)=0.082$
for $M_A=101$\,GeV.\footnote{
Here, we require the number of signal events to be less than 
two times the statistical fluctuation ($\sqrt{B}$) of the
SM background events. 
} We note that, as discussed in the previous
sections, this upper bound on $B(A \ra b \bar b) \times B(H^+ \ra
\nu_\tau \tau^+)$ as a function of $M_A$ (i.e., $m_{b \bar b}$) is
not sensitive to the other SUSY parameters in the MSSM.

Surely, the above bound can be improved by including the negatively charged
mode $AH^-$ of the signal event
(whose rate is about half of the $AH^+$
rate, cf. Fig.~\ref{fig:cross}) and the other decay mode
of $\tau^\pm$ (such as
$\tau^+ \ra \rho^+ {\bar \nu}_\tau$
and
$\tau^- \ra \rho^- {\nu_\tau}$).
Since the decay branching ratio B($\tau^+ \ra \rho^+ {\bar \nu}_\tau$)
is a factor of 2 larger than
B($\tau^+ \ra \pi^+ {\bar \nu}_\tau$), and the dominant
decay mode of $\rho^+ \ra \pi^+ \pi^0$ produces a $\pi^+$ whose typical
momentum is smaller than that produced in $\tau^+ \ra \pi^+ {\bar
\nu}_\tau$, we expect that including the
$\tau^+ \ra \rho^+ {\bar \nu}_\tau$ channel would increase the
event rate by about a factor of 2.
Therefore, after including both the $AH^+$ and $AH^-$
channels with $\tau$ decays into the $\pi \nu$ or $\rho \nu$ modes,
the signal rate will increase by about a factor of 3.
Assuming the background rates increase by the same factor,
the resolution power will increase by a factor about $\sqrt{3}$.
For example, the above constraint will read as
$B(A \ra b \bar b) \times B(H^\pm \ra \nu_\tau \tau^\pm) <
0.08/\sqrt{3}=0.046$ for $M_A=101$\,GeV in the MSSM.
However, a reliable conclusion can only be drawn
from the study using a full event generator (that predicts the
distributions of final state hadrons)
folded with realistic detector simulation.

Secondly, we consider Case B, in which the production rates of $HH^+$
and $AH^+$ are about the same.
Following the same analysis procedures as in Case A, we obtain the results
shown in Table~\ref{tbl:data3} and Fig.~\ref{fig:lsmear}.
According to the last column in Table~\ref{tbl:data3},
if a $AH^+$ signal is not found, then at the two standard deviation
level, $B(A \ra b \bar b) \times  B(H^+ \ra \nu_\tau \tau^+)$
in the MSSM has to be less than
$\displaystyle \frac{2 \sqrt{50}}{65} \times (0.90)(0.90)=0.18$
for $M_A=166$\,GeV.
After including the $AH^-$ production channel and
$\tau \ra \rho \nu$ decay modes, we expect the above bound to improve
by about a factor of $\sqrt{3}$, i.e.,
$B(A \ra b \bar b) \times B(H^\pm \ra \nu_\tau \tau^\pm) <
0.18/\sqrt{3}=0.1$ for $M_A=166$\,GeV in the MSSM.

\begin{table}
\caption{Numbers of $AH^+$ signal and background events for Case A,
with $M_A=101$\,GeV
in the $b {\bar b} \pi^+ {\slash E_T}$ channel, at the Tevatron with an
integrated luminosity of $30\,{\rm fb}^{-1}$. The $b$-tagging efficiency
($50\%$, for tagging both $b$ and ${\bar b}$ jets) is included, and
the kinematic cuts listed in each column are applied sequentially.}
\label{tbl:tevatron}
\begin{tabular}{c|c|c|c|c}
\hline
\hline
 &Basic Cuts & $\slash E_T>50$\,GeV & $p_T^{\pi}>40\,$GeV & $90<M_A<110$\,GeV\\
& & & &or\\
& & & &$105<M_H<125$\,GeV\\
\hline
$ AH^+$ & 11  & 8  & 3  & 3\\
$ HH^+$ & 1   & 1  & 0  & 0\\
\hline
$Wb\ov{b}$    & 383 & 27  & 3  & 0\\
$t\ov{b}$     & 6   & 3   & 0  & 0\\
$t\ov{t}$     & 0   & 0   & 0  & 0\\
$Wg$          & 0   & 0   & 0  & 0\\
\hline
\hline
\end{tabular}
\end{table}

Thirdly, we consider Case C, in which $H^+$ predominantly decays into
the $t \bar b$ mode. For the
CP-odd Higgs boson decays into a $b \bar b$ pair,
the dominant SM background comes from the
$b {\bar b} t {\bar t}$ production when
the decay products of $W^-$, produced from $\bar t$ decay,
escape detection.
The sensitivity of the LHC to this case is presented in Table~\ref{tbl:tb}
and Figs.~\ref{fig:tbnosmear} and \ref{fig:tbsmear}.
Due to the complexity of the background simulation, the result of our
parton level Monte Carlo analysis for the background event can only be
viewed as an estimate. A more reliable calculation using the full event
generator with detector simulation is needed to draw a definite
conclusion. Nevertheless, we illustrated that the
$t {\bar b}$ decay mode of the signal
event can be detected at the LHC with a reasonable resolution on the
determination of $M_A$ and $M_{H^+}$,
cf. Figs.~\ref{fig:tbnosmear} and \ref{fig:tbsmear}.
A crude estimate from Table~\ref{tbl:tb} reveals that
if a $AH^+$ signal is not found, then at the two standard deviation
level, $B(A \ra b \bar b) \times  B(H^+ \ra t {\bar b})$
has to be less than
$\displaystyle \frac{2 \sqrt{3}}{18} \times (0.89)(0.79)=0.135$
for $M_A=250$\,GeV in the MSSM.

Finally, we comment on the potential of the Fermilab Tevatron.
As shown in Fig.~\ref{fig:cross}, the production rate of the $AH^+$
event at the Tevatron
is only sizable for small $M_A$.
Assuming an integrated luminosity of $30\,{\rm fb}^{-1}$
at the Tevatron, the event yield for Case A is shown in
Table~\ref{tbl:tevatron}. It is evident that it is challenging to
detect such a signal
at the Tevatron.

\section{CONCLUSIONS}

\begin{figure}[t]
\begin{center}
\scalebox{0.4}{\includegraphics{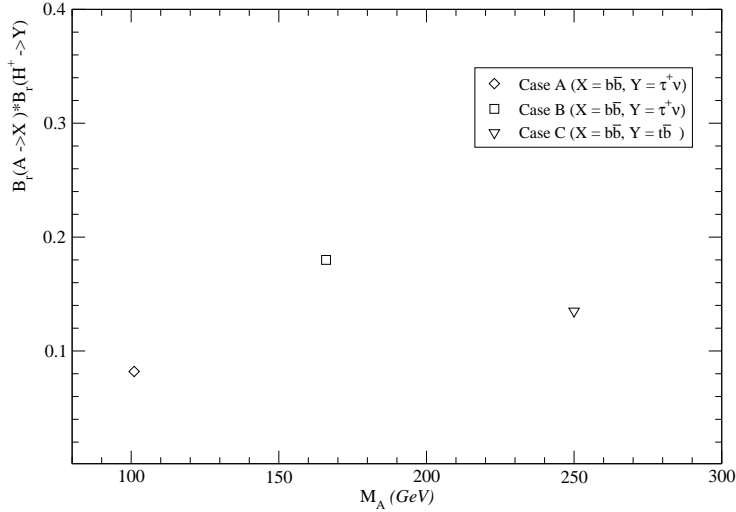}}
\end{center}
\vspace*{-5mm}
\caption{Constraints on the
product of branching ratios
$B(A \ra b {\bar b}) \times B(H^+ \ra \tau^+ \nu_\tau)$
as a function of $M_A$ for Case A and Case B,
 and
 $B(A \ra b {\bar b}) \times B(H^+ \ra t {\bar b})$
 for Case C, at the LHC, where
 $\tau^+$ decays into $\pi^+ {\bar \nu}_\tau$ channel.
 }
\label{fig:br-ma}
\end{figure}

Most production processes studied in the literature for testing the
MSSM contain at least two SUSY parameters (such as $\tan \beta$
and $M_A$) in the search for supersymmetric Higgs bosons.
Furthermore, the detection efficiency of the signal event depends on the
assumed decay channels of the SUSY particles, hence, on  the detailed
choice of SUSY parameters.
If the signal is not found after comparing experimental data with
theory prediction,
it is a common practice to constrain
the product of
the production cross section and the decay branching ratios
of final state SUSY particles as a function of
the multiple-dimension SUSY parameter space of the MSSM.

In Ref.~\cite{wah_plb}, a novel proposal was made to study the
$AH^\pm$ production process at hadron colliders.
It was pointed out that this process possesses the following interesting
properties:
\begin{itemize}
\item Its Born level production rate depends only on one
  SUSY parameter that can be determined by
  kinematic variables (e.g., the invariant mass of the $b \bar b $ pair).
\item Its higher order production rate is not sensitive to
  detailed SUSY parameters through electroweak radiative corrections.
\item Its final state particle kinematics can be properly modelled
  without specifying any SUSY parameters.
  Hence, the detection efficiency of the signal event
  can be accurately determined.
\item
  If the signal is found, it can be used to distinguish the MSSM
  Higgs sector from its alternatives,
  e.g.,  THDM.
\item
  If the signal is not found, one can
  constrain the MSSM by limiting the
  product of decay branching ratios alone,
  without convoluting with the production cross section.
\end{itemize}

At the LHC, the $AH^+$ signal event can be produced from the tree
level process $q \bar q' \to W^{+ \ast} \to A H^+$, whose rate
depends only on $M_A$ and $M_{H^+}$, for the coupling of
$W$-$A$-$H^+$ is fixed by the weak gauge coupling $g$, required by
$SU(2)$ gauge invariance. Furthermore, in the MSSM, $M_A$ and
$M_{H^+}$ are related as Eq.~(\ref{eq:massrel}), therefore the
$AH^+$ production rate depends only on one SUSY parameter, i.e.,
$M_A$, which can be kinematically determined from experimental
data by reconstructing the invariant mass of the decay particles
of $A$. For example, when $A$ decays into the $b \bar b$ channel,
the invariant mass of $b \bar b$ pair reveals the mass of $A$.
Since both $A$ and $H^+$ are scalar particles, the angular
distributions of their decay particles can be accurately modelled,
i.e., an isotropic distribution in its rest frame. If the signal
is found, the mass relation, Eq.~(\ref{eq:massrel}), can be
tested to distinguish the MSSM from the THDM
in which this mass relation does not generally hold. If the signal
event is not found, then one can constrain the MSSM parameters by
limiting the product of the decay branching ratios $B(A \ra X)
\times B(H^+ \ra Y)$ as a function of $M_A$, where $X$ and $Y$ are
the decay channels of $A$ and $H^+$, respectively, studied by
experimentalists.

In Sec.~II, we show that
the electroweak radiative corrections that
depend on detailed SUSY parameters are usually small 
in comparison to
the uncertainty in higher order (beyond the NLO) QCD corrections, 
parton distribution functions,
or the accuracy of the experimental measurement.

To test whether such a signal can be detected at the LHC, we
performed a Monte Carlo study at the parton level in Sec.~III.
We consider three cases of MSSM to cover the decay modes of
$A(\ra b {\bar b})H^+(\ra \tau^+ \nu_\tau)$
and
$A(\ra b {\bar b})H^+(\ra t {\bar b})$.
We concluded in the last part of Sec.~III that at the LHC
this signal event can indeed provide useful information about
the MSSM Higgs sector.
For example, if the $AH^+$ signal is not found in the decay mode of
$\tau^+ \ra \pi^+ {\bar \nu}_\tau$, then we can constrain the
product of branching ratios
$B(A \ra b {\bar b}) \times B(H^+ \ra \tau^+ \nu_\tau)$
as a function of $M_A$, as shown in Fig.~\ref{fig:br-ma}.
This corresponds to Case A or Case B defined in Table~\ref{tbl:susypara2}.
In case C, for $M_{H^+} > m_t + m_b$,
not finding the signal event implies an upper bound on
 $B(A \ra b {\bar b}) \times B(H^+ \ra t {\bar b})$
for a given $M_A$, and therefore constrains the 
MSSM as a function of one SUSY parameter. 
Including the negatively charged channel $AH^-$ and the
$\rho \nu$ decay mode of $\tau$ can tighten the above bounds roughly
by a factor of $\sqrt{3}$. However, to have a more accurate
conclusion, a full event generator with detector simulation
should be used to repeat the analysis outlined in this paper.

\begin{figure}[t]
\begin{center}
\scalebox{0.5}{\includegraphics{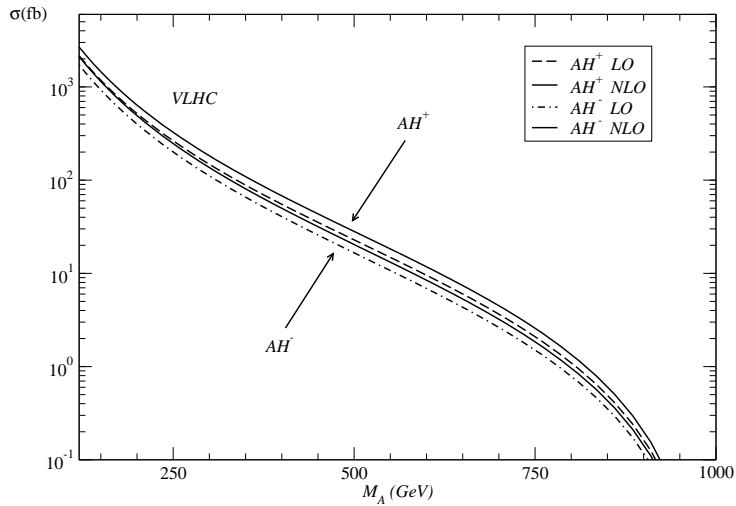}}
\end{center}
\vspace*{-5mm}
\caption{The LO (dotted lines) and NLO QCD (solid lines)
cross sections of the $AH^+$ and $AH^-$ pairs
as a function of $M_A^{}$ at the VLHC (a 200\,TeV $p p$ collider).}
\label{fig:cross2}
\end{figure}

From Fig.~\ref{fig:cross}, we see that the $AH^+$ rate becomes very small
(less than about $0.1$\,fb) at the LHC once $M_A$ is larger than
400\,GeV. Hence,
to cover the whole mass spectrum of the TeV scale MSSM, we
should find a high energy collider that can be sensitive to this process
for $M_A$ approaching the TeV region.
This could be one of the motivations for proposing a future
 Very Large Hadron Collider (VLHC), a 200\,TeV proton-proton collider.
 The signal rates at the VLHC are shown in Fig.~\ref{fig:cross2}.

\vspace*{5mm}
\noindent
{\bf Acknowledgments}~~~\\

We thank K.~Hagiwara, G.L.~Kane, H.K.~Kim, P.~Nadolsky,
Y.~Okada, T.~Tait and G. Tavares-Velasco for useful discussions.
This work was supported in part by the NSF grants PHY-0244919
and PHY-0100677.

\section{Appendices}
\subsection{Renormalization}

In this paper, we adopt the on-shell renormalization scheme
developed by Dabelstein\cite{dabelstein} to calculate the
one-loop electroweak corrections.
The standard model parameters are fixed by defining
$\alpha_{em}$, $m_W$ and $m_Z$, and the additional SUSY parameters
in the Higgs sector\footnote{
There are 7 parameters in the Higgs sector of the MSSM.
They are $g'$, $g$, $v_1$, $v_2$, $m_1$, $m_2$, and
$m_3$. Beyond the Born level, the wavefunction renormalization factors
$Z_{H_1}$ and $Z_{H_2}$ also need to be introduced to renormalize the
theory, where $H_1$ and $H_2$ denote the
two Higgs doublets in the model.
}
are fixed by the following renormalization
conditions:
(1) the tadpole contributions ($T_{H_1}=0$, $T_{H_2}=0$),
(2) the on-shell condition for the mass of $A$,
(3) the on-shell condition for the wavefunction of $A$,
(4) a renormalization condition on $\tan\beta$
   (which requires $\delta v_1/v_1=\delta v_2/v_2$), and
(5) a vanishing $A-Z$ mixing for an on-shell $A$.

\subsection{Calculation of $F^{(1)}(q^2)$}

The one-loop correction to
the renormalized form factor of the $W^\pm H^\mp A$ vertex,
apart from the effective weak gauge coupling ${\bar g}$,
can be written as

\begin{eqnarray}
F^{(1)}(q^2) &=& {\tilde Z}_{AA}^{1/2} {\tilde Z}_{H^+H^-}^{1/2} \left\{
1 + \delta F_{WHA}  \right.\nonumber\\
&&\left. + F_{WHA}^{\rm 1PI}(M_A^2,M_H^2,q^2) \right\}-1,
\label{eq:ffactor}
\end{eqnarray}
where ${\tilde Z}_{AA}$ and
${\tilde Z}_{H^+H^-}$ are the finite wavefunction
factors for the renormalization of the
external Higgs bosons $A$ and $H^\pm$.
In our scheme,
%
%\vspace*{-2mm}
%\noindent
\begin{eqnarray}
  {\tilde Z}_{AA} &=& 1,\\
  {\tilde Z}_{H^+H^-} &=& 1 - \Pi_{H^+H^-}'({M_A^2}+m_{W}^2)+
                  \Pi_{AA}'(M_A^2) \, ,
\end{eqnarray}
where $\Pi_{AA}'(M_A^2)$ denotes taking the derivative of the
two point function $\Pi_{AA}(k^2)$ of the CP-odd scalar $A$
with respect to $k^2$ at $k^2=M_A^2$, etc.
The terms inside the curly bracket of Eq.~(\ref{eq:ffactor})
arise from the renormalized vertex function of $WHA$.
$F_{WHA}^{\rm 1PI}(p_A^2,p_H^2, q^2)$ represents the one-loop
contribution of the
one-particle-irreducible (1PI) diagrams with $p_A^2,p_H^2,q^2$
as the four-momentum square of the
incoming $A$, $H^\mp$ and $W^\pm$ particles, respectively.
$\delta F_{WHA}$ is the counterterm contribution resulting
from the field renormalization of $H^+$ and $A$:
%
%\vspace*{-2mm}
%\noindent
\begin{eqnarray}
 H^+A  \to   H^+A \left(1+{1 \over 2}\delta Z_{H^+}+
 {1 \over 2}\delta Z_{A}\right).
\end{eqnarray}
In terms of the independent counterterms fixed by the
renormalization scheme, the wavefunction counterterms
$\delta Z_{H^+}$ and $\delta Z_{A}$
can be written as
$(\sin^2 \beta) \delta Z_{H_1} + (\cos^2 \beta) \delta Z_{H_2}$
which is found to be equal to
$-\frac{1}{2}\Pi_{AA}'(M_A^2)$.
We note that in $\delta F_{WHA}$ the contributions
from the counterterms of the weak gauge coupling and the
wavefunction renormalization of the $W$-boson
are not included, because they should be combined with the $W$-boson
self energy contribution to derive the
 running weak gauge coupling $\bar g (q^2)$.
In our numerical calculation, we use
$$
{\bar g}^2 = 4{\sqrt{2}}m_W^2G_F \, .
$$
In summary, the one-loop electroweak
correction to $F^{(1)}(q^2)$ is found to be
%
%\vspace*{-2mm}
%\noindent
\begin{eqnarray}
  &&F^{(1)}(q^2)  \equiv F_{WHA}^{\rm 1PI}(M_A^2,M_{H^\pm}^2,q^2)\nonumber\\
  &&- \frac{1}{2} \Pi_{H^+H^-}'(M_A^2+m_W^2)
  - \frac{1}{2} \Pi_{AA}'(M_A^2) .
\label{eq:ff}
\end{eqnarray}
In the above equation, the top- and bottom-loop
contribution to $F_{WHA}^{\rm 1PI}$ is given by
%
%\vspace*{-2mm}
\begin{eqnarray}
&&   F_{WHA}^{\rm 1PI (quark)}(q^2,p_A^2,p_H^2)   \nonumber\\
&&=\sum_{fff'=ttb,bbt} F_{WHA}^{fff'}(q^2,p_A^2,p_H^2),
\end{eqnarray}
%
%\vspace*{-2mm}
%\noindent
with
%
%\vspace*{-2mm}
\begin{eqnarray}
&&  F_{WHA}^{fff'}(p_A^2,p_H^2,q^2)
    = + \frac{N_c}{16\pi^2} y_f^2
\left\{ p_A^2 C_{31}^{fff'} - p_H^2 C_{32}^{fff'} \right.\nonumber\\
&&
         + (2 p_A \cdot p_H - p_A^2) C_{33}^{fff'}
         - (2 p_A \cdot p_H - p_H^2) C_{34}^{fff'} \nonumber \\
&&         + (D + 2) (C_{35}^{fff'} - C_{36}^{fff'})
         + p_A^2 C_{21}^{fff'} - (2 p_A \cdot p_H
\nonumber \\
&&
+ p_H^2) C_{22}^{fff'}
         - 2 p_A^2 C_{23}^{fff'}  - (D-2) C_{24}^{fff'}
         - m_f^2 C_{11}^{fff'}  \nonumber\\
&&
   \left.  - \left( q^2 + m_f^2 \right) C_{12}^{fff'} \right\}
       - c_f \frac{1}{16\pi^2} y_f y_{f'} m_f m_{f'} C_{0}^{fff'}.
\label{eq:qform1}
\end{eqnarray}
where $c_f=+1$ and $-1$ for $fff'=ttb$ and $bbt$, respectively, and
$C_{ij}^{fff'}$ are defined in terms of
the Passarino-Veltman functions\cite{PV} with
\begin{eqnarray}
     C_{ij}^{fff'} =
     C_{ij} \left(p_A^2, p_H^2, (p_A+p_H)^2; m_f,m_f,m_{f'}\right).
\end{eqnarray}
The stop- and sbottom-loop contribution is given by
%\vspace*{-2mm}
\begin{eqnarray}
&&  F_{WHA}^{\rm 1PI(squark)}
(p_A^2,p_H^2,q^2) = \frac{N_c}{16\pi^2} \sqrt{2} \sum_{i,j,k=1}^2
 \nonumber   \\
&&  \times  \left\{ U_{iL}^\ast D_{Lk}
    \left(i \lambda[\tilde{t}_j^\ast,\tilde{t}_i, A] \right)
    \lambda[\tilde{b}_k^\ast,\tilde{t}_j, H^-]
  \tilde{C}^{\tilde{t}_i\tilde{t}_j\tilde{b}_k} \right.\nonumber\\
&&  \left. -  U_{kL}^\ast D_{Li}
    \left(i \lambda[\tilde{b}_i^\ast,\tilde{b}_j, A] \right)
    \lambda[\tilde{b}_j^\ast,\tilde{t}_k, H^-]
  \tilde{C}^{\tilde{b}_i\tilde{b}_j\tilde{t}_k} \right\},
\label{eq:sqform1}
\end{eqnarray}
%
%\vspace*{-2mm}
%\noindent
where $U_{Ii}$, $D_{Ii}$ are the rotation
matrices for stops and sbottoms between
the weak eigenstate basis ($I=L,R$) and the mass eigenstate basis $(i=1,2)$,
respectively.
$\lambda[\tilde{f}_i^\ast,\tilde{f}_j', \phi_k]$ represents the
coefficient of the $\tilde{f}_i^\ast\tilde{f}_j' \phi_k$ interaction
in the MSSM Lagrangian, as listed in Appendix~{\bf C}, and
%\vspace*{-2mm}
%\noindent
\begin{eqnarray}
 \tilde{C}^{\tilde{f}_i\tilde{f}_j'\tilde{f}_k''}
= \left( C_{11} - C_{12} \right) \left(p_1^2,p_2^2,q^2;
     m_{\tilde{f}_i}, m_{\tilde{f}_j'}, m_{\tilde{f}_k''} \right).
\end{eqnarray}
The quark (top and bottom) and squark (stop and sbottom)
loop contributions to the self-energies
$\Pi_{AA}(q^2)$ and $\Pi_{H^+H^-}(q^2)$ can be found
in Appendix {\bf E}.

\subsection{Squark couplings with $H^\pm$ and $A$}

The mass eigenstates of the squarks
are related to their weak eigenstates by the rotation matrix
$O^{f\dagger}_{i\,I}$ with
$\tilde{f}_i = \sum_{I} O^{f\dagger}_{i\,I} \tilde{f}_I$,
where  $i=1,2$ and $I=L,R$; $O^{f}_{I\,i}=U_{I\,i}$ and $D_{I\,i}$
for $f=t$ and $b$, respectively. In terms of the mixing angles
$\theta_f$, we have
\begin{eqnarray}
 \left(\begin{array}{c} \tilde{f}_1 \\ \tilde{f}_2 \end{array}\right)
=\left( \begin{array}{cc} \cos \theta_f & \sin \theta_f \\
                    -\sin \theta_f & \cos \theta_f \\
  \end{array}\right)
 \left( \begin{array}{c} \tilde{f}_L \\ \tilde{f}_R \end{array} \right).
\end{eqnarray}
Here, we define the mixing angle $\theta_f$ so that
$\tilde{f}_1$ is lighter than $\tilde{f}_2$.

The coupling constants among the weak-eigenstate squarks
and the Higgs bosons are defined through the Lagrangian
\begin{eqnarray}
  {\cal L} = \cdot\cdot\cdot
 +  \lambda[\tilde{f}_I^\ast,\tilde{f}'_J,\phi,...]
                           \tilde{f}_I^\ast \tilde{f}'_J \phi,...
+ \cdot\cdot\cdot.
\end{eqnarray}
Hence, the coupling constants for the
mass-eigenstate squarks
are a linear combination of the couplings for
the weak-eigenstate squarks, and
\begin{eqnarray}
\lambda[\tilde{f}_i^\ast,\tilde{f}'_j,\phi,...]
 &=&
\lambda[
\tilde{f}_I^\ast O^{f}_{I\,i}, O^{f'\dagger}_{i\,I} \tilde{f}'_J,\phi,...]
\nonumber \\
  &=&O^{f}_{I\,i} O^{f'\dagger}_{i\,I}
\lambda[\tilde{f}_I^\ast,\tilde{f}'_J,\phi,...].
\end{eqnarray}
The relevant couplings
$\lambda[\tilde{f}_I^\ast,\tilde{f}'_J,\phi,...]$,
denoted as
$\lambda_{\tilde{f}_I^\ast \tilde{f}'_J \phi,...}$,
are listed below.
%
%\vspace*{-2mm}
\begin{eqnarray}
&&\lambda_{\tilde{b}_L^\ast \tilde{t}_L H^-} =
\frac{-\sqrt{2}}{v}
(m_W^2 \sin 2 \beta - m_b^2 \tan\beta - m_t^2 \cot\beta) ,\\
&&\lambda_{\tilde{b}_L^\ast \tilde{t}_R H^-} =
\frac{\sqrt{2}m_t}{v}(A_t \cot\beta+\mu)   \label{coup1},\\
&&\lambda_{\tilde{b}_R^\ast \tilde{t}_L H^-} =
\frac{\sqrt{2}m_b}{v}(A_b \tan\beta+\mu)   \label{coup2},\\
&&\lambda_{\tilde{b}_R^\ast \tilde{t}_R H^-} =
\frac{2\sqrt{2} m_t m_b}{v \sin 2\beta},\\
&&\lambda_{\tilde{f}_L^\ast \tilde{f}_L A} =
\lambda_{\tilde{f}_R^\ast \tilde{f}_R A} =  0,
(\tilde{f}=\tilde{t},\tilde{b}),\\
&&\lambda_{\tilde{b}_L^\ast \tilde{b}_R A} =
\frac{i m_b}{v}(A_b \tan\beta+\mu) \label{coup3},\\
&&\lambda_{\tilde{t}_L^\ast \tilde{t}_R A} =
\frac{i m_t}{v}(A_t \cot\beta+\mu) \label{coup4},\\
&&\lambda_{\tilde{b}_L^\ast \tilde{b}_L A A} =
\frac{-m_b^2}{v^2} \tan^2\beta + \frac{g_Z^2}{4}(T_b-Q_b {s}_W^2)
\cos2\beta,\\
&&\lambda_{\tilde{b}_R^\ast \tilde{b}_R A A} =
\frac{-m_b^2}{v^2} \tan^2\beta + \frac{g_Z^2}{4}Q_b {s}_W^2
\cos2\beta  ,\\
&&\lambda_{\tilde{b}_L^\ast \tilde{b}_L A A} =
\frac{-m_t^2}{v^2} \cot^2\beta + \frac{g_Z^2}{4}(T_t-Q_t {s}_W^2)
\cos2\beta,\\
&&\lambda_{\tilde{b}_R^\ast \tilde{b}_R A A} =
\frac{-m_t^2}{v^2} \cot^2\beta + \frac{g_Z^2}{4}Q_t {s}_W^2
\cos2\beta  ,\\
&&\lambda_{\tilde{b}_L^\ast \tilde{b}_L H^+H^-} =
\frac{-2 m_t^2}{v^2} \tan^2\beta + \frac{g_Z^2}{2}(T_t-Q_t {s}_W^2)
\cos2\beta,\\
&&\lambda_{\tilde{b}_R^\ast \tilde{b}_R H^+H^-} =
\frac{-2 m_b^2}{v^2} \tan^2\beta + \frac{g_Z^2}{2}Q_b {s}_W^2
\cos2\beta  ,\\
&&\lambda_{\tilde{t}_L^\ast \tilde{t}_L H^+H^-} =
\frac{-2m_b^2}{v^2} \tan^2\beta + \frac{g_Z^2}{2}(T_b-Q_t {s}_W^2)
\cos2\beta,\\
&&\lambda_{\tilde{t}_R^\ast \tilde{t}_R H^+H^-} =
\frac{-2m_t^2}{v^2} \cot^2\beta + \frac{g_Z^2}{2}Q_t {s}_W^2
\cos2\beta,
\end{eqnarray}
where $T_t,T_b,Q_t$ and $Q_b$ are $\frac{1}{2},\frac{-1}{2},\frac{2}{3}$
and $\frac{-1}{3}$, respectively, and
\begin{eqnarray}
&&\lambda_{\tilde{t}_I^\ast \tilde{b}_J H^+} =
 \lambda_{\tilde{b}_I^\ast \tilde{t}_J H^-},\\
&&\lambda_{\tilde{f}_R^\ast \tilde{f}_L A} =
 -\lambda_{\tilde{f}_L^\ast \tilde{f}_R A},
\end{eqnarray}
for $I,J=L,R$ and $\tilde{f}=\tilde{t}, \tilde{b}$.

\subsection{Squark contributions to the $\rho$ parameter}
%{\bf Squark contributions to the $\rho$ parameter}

The squark one-loop contribution
to the $\rho$ parameter is given by
%
%\vspace*{-2mm}
\begin{eqnarray}
  \Delta \rho = \rho -1 = - 4 \sqrt{2} G_F
       {\rm Re}[ \Delta \Pi^{11}_T(0) - \Delta \Pi^{33}_T(0) ],
\end{eqnarray}
%
%\vspace*{-2mm}
%\noindent
with~\cite{pifunc}
%
%\vspace*{-2mm}
\begin{eqnarray}
      \Delta \Pi^{11}_T(q^2) &=& \frac{N_c}{16\pi^2}
                       \sum_{f=t,b}  \sum_{i,j=1}^2
                       {T_{f_L}}^2 |O_{Li}^f|^2 |O_{Lj}^f|^2 \nonumber\\
 &&  \;\;\;\;\;\;\;\;\;\;\;\;\;\;\;\;\;\;\;\;\;\;\;\;\;
                     \times B(q^2;m_{\tilde{f}_i}^2,m_{\tilde{f}_j}^2),\\
      \Delta \Pi^{33}_T(q^2) &=& \frac{N_c}{32\pi^2}
                       \sum_{i,j=1}^2 |U_{Li}|^2 |D_{Lj}^f|^2
                       B(q^2;m_{\tilde{u}_i}^2,m_{\tilde{d}_j}^2),
\end{eqnarray}
%
%\vspace*{-2mm}
%\noindent
where $O_{Ii}^t=U_{Ii}$ and $O_{Ii}^b=D_{Ii}$;
$B(q^2;m_1^2,m_2^2) \equiv
A(m_1^2)+A(m_2^2)-4 B_{22}(q^2;m_1^2,m_2^2)$.
By using the expression
\begin{eqnarray}
B(0;m_1^2,m_2^2)=-\frac{1}{2}(m_1^2+m_2^2)
+\frac{m_1^2m_2^2}{m_1^2-m_2^2} \ln\frac{m_2^2}{m_1^2} \, .
\end{eqnarray}
Eq.~(\ref{eq:rho}) is deduced under the assumption that
$M^2=M_Q^2 \simeq M_U^2 \simeq M_D^2 \gg m_t^2$ and
the stop mixing is large
($m_t|X_t|\simeq M^2$ and $m_b|X_b| \simeq 0$),
so that $m_{{\tilde t}_1} \sim {\cal O}(m_Z)$, which yields
$m_{{\tilde t}_2} \sim \sqrt{2} M$,
and $m_{{\tilde b}_1} \sim m_{{\tilde b}_2} \sim M$.

\subsection{Self energies}

%quark
The (top and bottom) quark-loop contributions to the self-energies
$\Pi_{AA}(q^2)$ and $\Pi_{H^+H^-}(q^2)$ in $D$-dimensions 
are expressed
in terms of the Passarino-Veltman functions~\cite{PV} by
%
%\vspace*{-2mm}
\begin{eqnarray}
&&  \Pi_{AA}^{\rm quark} (q^2) = - \frac{N_c}{16\pi^2}\,
    \sum_{f=t,b}
   2  y_f^2 \left\{    q^2 \left( B_1(q^2,m_f,m_f)
\right.\right. \nonumber\\
&&   \left.                + B_{21}(q^2,m_f,m_f) \right)
                     + D B_{22}(q^2,m_f,m_f)\nonumber \\
&& \left.
                     + m_f^2  B_{0}(q^2,m_f,m_f)
                                \right\}, \label{eq:qform2} \\
&&  \Pi_{H^+H^-}^{\rm quark} (q^2) =
  - \frac{N_c}{16\pi^2}\,2\,(y_b^2 + y_t^2)
       \left\{       q^2  \left( B_1(q^2,m_b,m_t)
\right.\right.
\nonumber\\
&&  \left.\left.
                           +  B_{21}(q^2,m_b,m_t) \right)
                     + D  B_{22}(q^2,m_b,m_t) \right\}, \nonumber\\
&&  - \frac{N_c}{16\pi^2}\,4\,y_b y_t m_b m_t
B_{0}(q^2,m_b,m_t).  \label{eq:qform3}
\end{eqnarray}

%squark
The stop- and sbottom-loop contributions are given by
%
%\vspace*{-2mm}
\begin{eqnarray}
&&\Pi_{AA}^{\rm squark} (q^2) =
- \frac{N_c}{16\pi^2} \sum_{\tilde{f}=\tilde{t},\tilde{b}}\;
     \sum_{i,j=1}^2   \nonumber\\
&&\times \lambda[\tilde{f}_i^\ast,\tilde{f}_j, A]
       \lambda[\tilde{f}_j^\ast,\tilde{f}_i, A]
       B_0(q^2,m_{\tilde{f}_i},m_{\tilde{f}_j})
\nonumber\\ 
&& - \frac{N_c}{16\pi^2} 2 \sum_{\tilde{f}=\tilde{t},\tilde{b}}\;
     \sum_{i=1}^2 \lambda[\tilde{f}_i^\ast,\tilde{f}_i,A,A]
       A(m_{\tilde{f}_i}),  \label{eq:sqform2}\\
&&\Pi_{H^+H^-}^{\rm squark} (q^2) = - \frac{N_c}{16\pi^2}
       \sum_{i,j=1}^2  \nonumber\\
&&       \times \lambda[\tilde{t}_i^\ast,\tilde{b}_j, H^+]
       \lambda[\tilde{b}_j^\ast,\tilde{t}_i, H^-]
       B_0(q^2,m_{\tilde{t}_i},m_{\tilde{b}_j})
\nonumber\\ 
&& - \frac{N_c}{16\pi^2} \sum_{\tilde{f}=\tilde{t},\tilde{b}}\;
     \sum_{i=1}^2 \lambda[\tilde{f}_i^\ast,\tilde{f}_i,H^+,H^-]
       A(m_{\tilde{f}_i})\, .  \label{eq:sqform3}
\end{eqnarray}
The self-energy $\Pi_{WW}^{}(q^2)$ of the $W$ boson was already presented
in the literature. For example, the quark-loop contribution can be
found in Ref.~\cite{hhkm}, and the squark-loop contribution in
Ref.~\cite{pifunc}.

%\end{narrowtext}
\end{document}